\documentclass{elsarticle}
\usepackage[T1]{fontenc}
\usepackage[utf8]{inputenc}
\usepackage{lmodern}
\usepackage{amsmath,amssymb,mathtools,bm}
\usepackage{graphicx,booktabs,array,longtable,placeins}
\usepackage[export]{adjustbox}
\usepackage[dvipsnames]{xcolor}
\usepackage[colorlinks=true,linkcolor=MidnightBlue,citecolor=MidnightBlue,urlcolor=MidnightBlue]{hyperref}
\usepackage{caption}
\usepackage[margin=1in]{geometry}

\biboptions{sort&compress}
\newcommand{\CP}{\mathbb{C}\mathrm{P}}
\newcommand{\Torus}{\mathbb{T}}
\newcommand{\Log}{\operatorname{Log}}
\newcommand{\Exp}{\operatorname{Exp}}
\newcommand{\PT}{\mathcal{P}}
\newcommand{\FS}{\mathrm{FS}}
\newcommand{\SPD}{\mathrm{SPD}}
\newcommand{\diag}{\operatorname{diag}}
\newcommand{\Reals}{\mathbb{R}}
\newcommand{\E}{\mathbb{E}}
\newcommand{\tr}{\operatorname{tr}}
\newcommand{\fittable}[1]{\adjustbox{max width=\textwidth}{#1}}
\hypersetup{pdftitle={Geometry, masking and selection in Differential Evolution and SOMA on quantum state manifolds},pdfauthor={Vojtech Novak, Ivan Zelinka}}
\journal{Swarm and Evolutionary Computation}

\begin{document}

\begin{frontmatter}

\title{Geometry, masking and selection in Differential Evolution and SOMA on quantum state manifolds}

\author[vsb,it4i,ku]{Vojt\v{e}ch Nov\'ak\corref{cor1}}
\ead{vojtech.novak.st1@vsb.cz}
\author[vsb,it4i,ku]{Ivan Zelinka}
\cortext[cor1]{Corresponding author}

\affiliation[vsb]{organization={Department of Computer Science, VSB--Technical University of Ostrava},
            city={Ostrava},
            country={Czech Republic}}
\affiliation[it4i]{organization={IT4Innovations National Supercomputing Center},
            city={Ostrava},
            country={Czech Republic}}
\affiliation[ku]{organization={Marine Research Institute, Klaipėda University},
            city={Klaipėda},
            country={Lithuania}}

\begin{abstract}
Differential Evolution (DE) and the Self-Organizing Migrating Algorithm (SOMA) generate new candidates through vector arithmetic: they subtract population members, scale the difference, and mask coordinates. Many practical search spaces are not vector spaces, however. Rotation angles are periodic, normalized states lie on spheres and projective spaces, and circuit parameters are coupled through the quantum states they produce. We present a systematic study of how such geometry affects the core operations of DE and SOMA. Keeping the operators and selection rules fixed, we replace only their geometric primitives and separate the effects of difference construction, masking, and selection in a large paired numerical study covering textbook manifolds and variational quantum problems. Geometry-consistent differences make full-crossover ($CR=1$) DE significantly better on 16 of 31 non-Euclidean configurations, with no significant deterioration, and the advantage persists when the measured computational cost of each variant is taken into account. The effect is largest where ordinary subtraction misrepresents distances between candidates: success on periodic synchronization rises from 44\% to 100\%, and on a positive-definite matrix problem from 1\% to 99\%. On six-qubit variational eigensolvers, shortest-path angular differences reduce the mean final loss by 90\% for about 3\% extra runtime. The improvements come with two contrasting diversity patterns, more effective population contraction or preserved population diversity, so higher acceptance rates do not necessarily indicate better search. Crossover masks interact with geometry: they can favor cheaper retractions and can increase the distance moved in quantum-state space when circuit parameters are coupled. These results give practical rules for designing population-based optimizers on non-Euclidean and quantum search spaces.
\end{abstract}


\end{frontmatter}

\section{Introduction}

Differential Evolution (DE) proposes a new candidate through the familiar rule $a+F(b-c)$: read a direction from two donors, attach it to a third individual, and scale the step~\cite{storn1997,das2011,das2016}. SOMA follows a related principle, sampling masked steps from an individual toward the current leader~\cite{zelinka2004,soma2016,skanderova2023}. These operations are straightforward in a vector space. They become ambiguous when two parameter values represent the same point, or when feasible solutions lie on a curved space. For example, angles near zero and $2\pi$ are neighbors, yet their raw subtraction is almost a full turn. On a sphere, an ordinary vector sum generally leaves the feasible surface.

\begin{figure}[!htbp]
\centering\includegraphics[width=\textwidth]{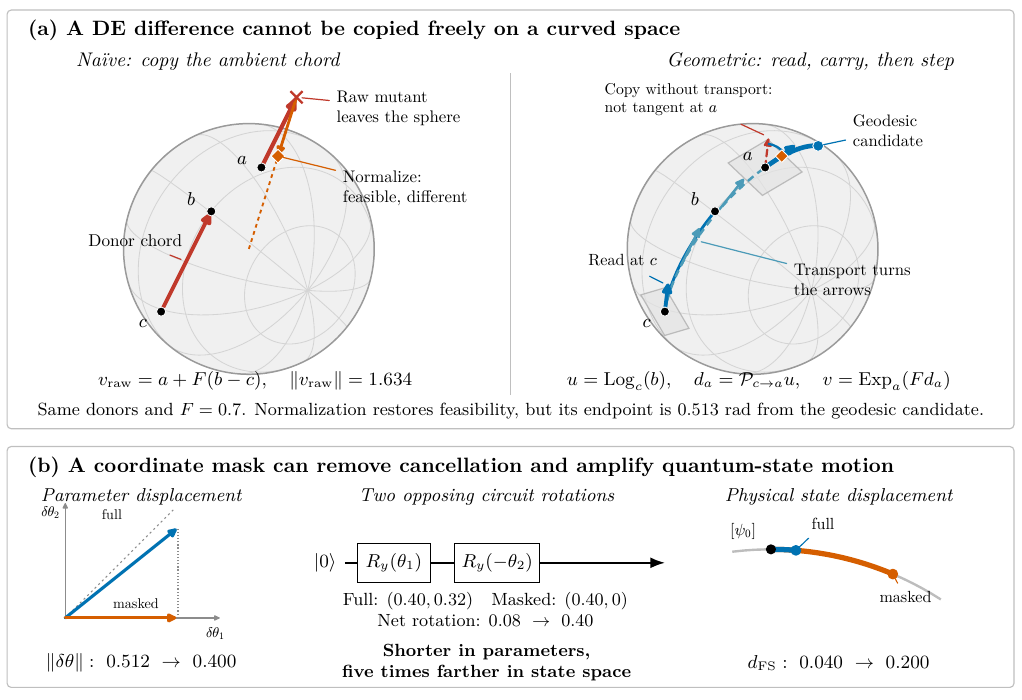}
\caption{Two geometric effects on population proposals. (a) Identical donors produce different sphere updates. Raw DE leaves the surface; normalization does not recover the geometric candidate. Even the donor logarithm loses tangency if copied to $a$ (red dashed arrow). Transport keeps it tangent along the path, preserving length while changing its ambient orientation; Exp follows the surface. The benchmark's feasible ambient variant projects and retracts, rather than simply normalizing raw DE. (b) Removing one of two opposing circuit rotations shortens the parameter step yet increases Fubini--Study state distance fivefold. Arcs are magnified. These are worked examples without objective ordering; the sphere is not a general embedding of quantum state space.}
\label{fig:overview}
\end{figure}

The central question is operational: \emph{which familiar properties of DE and SOMA survive these changes, and does a more exact geometric update produce better selected solutions for its cost?} Figure~\ref{fig:overview} summarizes the study. The changes are applied at the proposal level, while the original selection rules are retained. This separates a geometric property, such as preserving the length of a transported direction, from an optimization outcome, such as finding a lower loss. The distinction matters because respecting the geometry of a search space guarantees neither effective exploration nor convergence. An evolutionary algorithm depends not only on how candidates are generated but also on how selection lets them change the population, so a geometrically exact update can lose to a simpler approximation if it produces unfavorable step sizes or interacts poorly with crossover. Understanding these interactions is the central motivation of this study.

A manifold can be understood here as a feasible search space that looks like a vector space in a sufficiently small neighborhood. Its tangent space is the local workspace for directions. Established manifold optimization supplies three tools: a logarithmic map reads a displacement between points, transport carries a direction to another base, and an exponential map or retraction converts a tangent step into a feasible candidate~\cite{absil2008,boumal2023,bonnabel2013}. These operations already underpin evolutionary and swarm search on matrix manifolds~\cite{colutto2010,arnold2014,borckmans2010,he2020,shan2026}, intrinsic DE mutation built from discrete geodesics on meshes~\cite{liu2016mde}, and projection-based DE on the Grassmann manifold~\cite{lesniewski2025}. How to run DE on a manifold is therefore not the contribution of this paper. Its contribution is a controlled investigation of how the choice of geometric operation interacts with crossover, population statistics, and selection when everything else is held fixed.

This perspective connects directly to evolutionary computation theory. DE donor differences inherit information from the population covariance~\cite{opara2018,opara2019}; SOMA's Euclidean proposals admit exact masking and leader-distance identities~\cite{novak2026,novak2026isoma}. Crossover changes both direction and active dimensionality, which makes DE coordinate dependent~\cite{zaharie2009,sutton2007,hansen2011}. Recent large-scale DE comparisons also emphasize that favorable choices depend on problem class and budget~\cite{piotrowski2026}. A useful geometric benchmark should therefore explain changes in these mechanisms and their costs, rather than report feasibility or an overall ranking alone.

Variational quantum algorithms (VQAs) provide structured test problems for this purpose~\cite{cerezo2021}. A quantum circuit maps adjustable angles to a normalized state vector, and a measured observable or prediction supplies the loss. Global phase does not change the physical state, so equivalent state vectors are identified as one point in complex projective space. This identification is a \emph{quotient}: several numerical descriptions represent the same physical solution. A variational quantum eigensolver (VQE) minimizes energy~\cite{peruzzo2014}; the quantum approximate optimization algorithm (QAOA) targets a combinatorial objective~\cite{farhi2014}; quantum regression minimizes prediction error. These examples combine periodic controls, redundant descriptions, and coupled effects of parameters. Quantum natural gradient methods use the local metric induced by the circuit~\cite{stokes2020}; previous work also applies DE directly to variational circuits~\cite{failde2023,novak2026qaoa}, and classical optimizers for variational algorithms have been compared extensively~\cite{lavrijsen2020,bonetmonroig2023,novak2026geom}. Our question concerns finite population displacements and masks, which are not determined by a local gradient alone.

The benchmark has three linked parts. First, geometry-only controls test the flat covariance and SOMA contraction laws from local to dispersed populations. Second, paired geometry and mask ablations compare final loss across 33 textbook and quantum configurations. Third, proposal diagnostics and recorded run times connect those outcomes to selection and computational cost. We distinguish the flat parameter domain from its curved state image throughout; an unrestricted state-space step is used only where that state family is feasible.

The study makes four contributions.
\begin{enumerate}
\item It separates the effects of the geometric operation, the mask, and selection in fixed DE and SOMA operators, using paired initialization over 33 textbook and quantum configurations.
\item It explains, in algorithmic terms, three mechanisms that predict where the Euclidean intuition fails: when DE differences stop tracking the population spread, when a coordinate mask enlarges rather than shrinks the change of the quantum state, and why exact full-mask SOMA can collapse onto its leader (Section~\ref{sec:props}).
\item It reports the outcomes at matched evaluations, at matched wall time, and at a fixed target. Full-crossover ($CR=1$) DE benefits from shortest-path geometry in 16 of 31 non-Euclidean configurations; the conclusion holds when measured run times are taken into account, while success probabilities reveal order-of-magnitude differences that mean losses understate.
\item It identifies two contrasting diversity patterns associated with the gains, and cases where geometry does not help: partial crossover can favor a cheaper retraction, and projective quantum problems are largely insensitive to the primitive.
\end{enumerate}

\section{Methods}

\subsection{Search geometry and the quantum circuit map}

\begin{figure}[!htbp]
\centering\includegraphics[width=.88\textwidth]{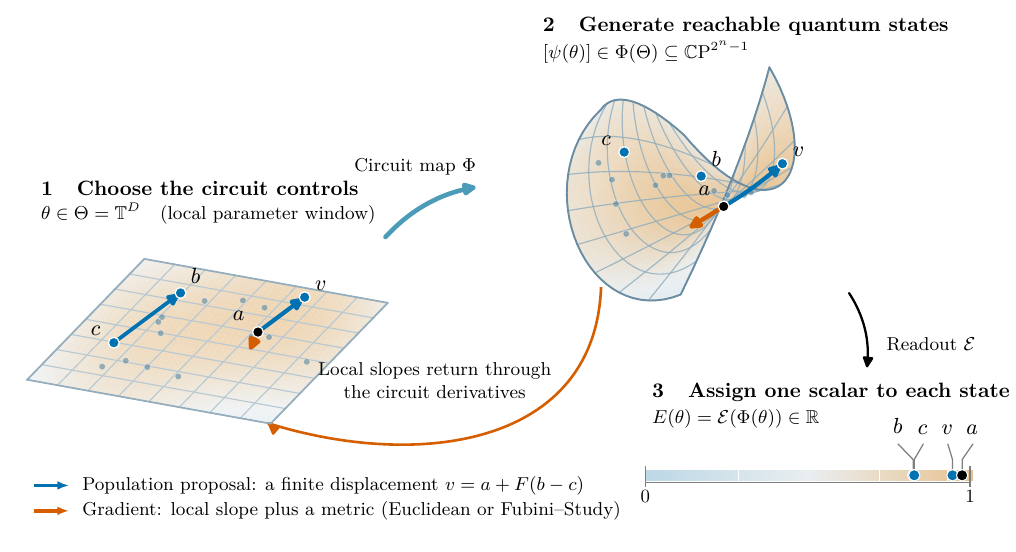}
\caption{One population, three representations. A local grid of periodic controls maps to a curved image of reachable states, then to scalar energies. Matching labels follow the same candidates; colour indicates energy on both the parameter grid and state image. The circuit is $\operatorname{CR}_y(\theta_1)(R_y(\theta_1)\otimes R_y(\theta_2))|00\rangle$. Its real state rays are displayed in a local three-dimensional chart, not an isometric embedding of $\CP^3$. Blue arrows show the finite proposal $v=a+0.7(b-c)$ and its circuit image. Orange arrows show Euclidean parameter descent and FS descent restricted to the reachable image; the return curve denotes derivative information. This is a computed illustration, not an additional optimizer comparison.}
\label{fig:domainmap}
\end{figure}

Figure~\ref{fig:domainmap} distinguishes the spaces hidden inside a quantum cost evaluation: the parameters manipulated by the optimizer, the quantum states they produce, and the scalar values used by selection. Distances in one space need not correspond to distances in another. Two nearby parameter vectors can produce very different states, and two distant ones can produce the same state, so the size of an optimizer's parameter update does not describe the size of its physical effect. For an energy objective, the computation is
\begin{equation}
 \theta\in\Theta=\Torus^D
 \ \xrightarrow{\ \Phi\ }\ [\psi(\theta)]\in\mathcal A=\Phi(\Theta)\subseteq\CP^{2^n-1}
 \ \xrightarrow{\ \mathcal E\ }\ E(\theta)\in\Reals,
 \qquad E=\mathcal E\circ\Phi.
 \label{eq:three_spaces}
\end{equation}
The first space contains adjustable controls. The circuit map $\Phi$ generates the reachable state image $\mathcal A$ inside its state-space codomain; $\mathcal A$ need not fill that codomain. The readout $\mathcal E([\psi])=\langle\psi|H|\psi\rangle$ then assigns one real energy. Thus, the state image is not the scalar range of the cost function. It is generally a nonlinear set of physical rays, rather than a linear Hilbert subspace. Figure~\ref{fig:domainmap} displays a two-qubit real-state family in a local three-dimensional coordinate chart. Its drawn shape and distances are coordinate-dependent, rather than an isometric picture of the Fubini--Study (FS) state geometry defined below. Regression additionally combines circuit outputs into a training loss.

The distinction between population and gradient methods concerns how directions are constructed, not exclusive use of different spaces. A donor difference is a finite displacement between two candidates. A gradient is a local slope, $dE_\theta[u]=d\mathcal E_{\Phi(\theta)}[d\Phi_\theta(u)]$ by the chain rule, which a metric converts into a direction~\cite{boumal2023}. Ordinary gradient descent therefore steps in parameter space, as angular DE does, whereas quantum natural gradient uses the circuit-induced metric to construct parameter updates~\cite{stokes2020}. Intrinsic population methods may likewise construct differences in a known feasible state manifold.

The practical meaning of the geometric operations is summarized in Table~\ref{tab:translation}. Unlike a gradient direction, which is computed at one current point, a DE donor displacement is obtained from two points and used at a third. It must therefore specify both where it is read and how it is moved. A SOMA leader displacement is already based at its migrant, so it does not require this between-donor transport.

\begin{table}[!htbp]
\centering\small
\caption{The ordinary operations retained by DE and SOMA, and their geometric interpretation. These are standard tools; geometric exactness refers to the chosen metric, not to objective improvement.}
\label{tab:translation}
\begin{tabular}{p{25mm}p{36mm}p{86mm}}
\toprule
Operation & Flat-space interpretation & Manifold interpretation\\
\midrule
Difference & $b-c$ & $\Log_c(b)$: read the initial direction and distance of a shortest path from $c$ to $b$.\\
Carry a direction & Attach the same vector at $a$ & $\PT_{c\to a}$: move it along a declared path while preserving its metric length; the resulting orientation can depend on the path.\\
Make a candidate & $a+u$ & $\Exp_a(u)$: follow the geodesic with initial velocity $u$; or use a retraction $R_a(u)$ that agrees to first order near zero.\\
Mask coordinates & Set components to zero & Choose a local frame, mask its coefficients, and reconstruct the tangent direction.\\
Select & Compare objective values & Retain DE target replacement or SOMA best-path selection, separately from how the points were generated.\\
\bottomrule
\end{tabular}
\end{table}

Let $(M,g)$ denote the declared search domain and its rule for measuring lengths. On a flat torus with coordinate periods $L_j$, the shortest logarithmic displacement and exponential map are
\begin{equation}
 [\Log_x(y)]_j=((y_j-x_j+L_j/2)\bmod L_j)-L_j/2,
 \qquad [\Exp_x(u)]_j=(x_j+u_j)\bmod L_j.
 \label{eq:torus}
\end{equation}
Parallel transport is the identity in these flat angular coordinates. Wrapping an endpoint alone does not make raw donor subtraction a shortest-path difference. The torus has nontrivial periodic topology but zero intrinsic curvature: its global seam ambiguity must be distinguished from transport effects on curved domains.

For a unit representative $p\in\mathbb{C}^{m}$, complex projective space $\CP^{m-1}$ identifies $p$ with $e^{i\alpha}p$. We use
\begin{equation}
 d_\FS([p],[q])=\arccos|p^\dagger q|,\qquad
 T_{[p]}\CP^{m-1}=\{u:p^\dagger u=0\},\qquad
 g_p(u,v)=\operatorname{Re}(u^\dagger v).
 \label{eq:fs}
\end{equation}
The tangent representation is horizontal: the physically irrelevant phase direction $ip$ is excluded. FS distance is the physical state-distance convention used throughout. This normalization has pure-state quantum Fisher information matrix $Q=4g$. Projective representatives are phase-aligned before logarithms or ambient chords are formed~\cite{manifoldsjl}. The sphere uses its induced metric, the product $(\CP^1)^n$ uses the product metric, and symmetric positive-definite (SPD) matrices use the affine-invariant metric. Implementation formulas and the curvature properties of these domains are given in \ref{app:geometry}.

Parallel transport around a loop can return a vector with a changed orientation, an effect called \emph{holonomy}; we use it to validate the implementation. At a \emph{cut locus}, such as a sphere antipode, several shortest paths can compete, so a shortest logarithm needs a branch convention.

A circuit defines a map $\Phi:\Torus^D\to\CP^{2^n-1}$, not an identification of its parameter domain with the state tangent space. Its pullback metric is
\begin{equation}
 g^\Phi_{ij}(\theta)=\operatorname{Re}\left(
 \langle\partial_i\psi|\partial_j\psi\rangle-
 \langle\partial_i\psi|\psi\rangle\langle\psi|\partial_j\psi\rangle\right).
 \label{eq:pullback}
\end{equation}
This matrix can be singular, and the reachable set need not be a globally smooth embedded manifold. General layered circuits are therefore optimized in their parameter tori. Intrinsic state-space updates are restricted to models with explicitly known feasible manifolds: unrestricted states, product states, fixed-unitary images of product states, and the invariant projective line of single-edge one-layer QAOA. An unrestricted $\CP$ step is never substituted for a feasible update of a general circuit.

The cancellation example in Fig.~\ref{fig:overview}(b) makes this distinction explicit. Its two rotations combine into the net angle $\theta_1-\theta_2$:
\begin{equation}
 |\psi(\theta)\rangle=R_y(-\theta_2)R_y(\theta_1)|0\rangle,
 \qquad g^\Phi=\tfrac14\begin{bmatrix}1&-1\\-1&1\end{bmatrix}.
 \label{eq:cancellation}
\end{equation}
From $(0,0)$, the full step $(0.40,0.32)$ has parameter norm 0.512 and FS distance 0.040; masking the second coordinate gives norm 0.400 but FS distance 0.200. The pullback has rank one. The larger state change comes from removing cancellation, not from increasing parameter step length.

\subsection{Fixed DE and SOMA operators}

For target $x_i$, three distinct donors $a,b,c$, also distinct from the target, define the intrinsic DE/rand/1 mutant
\begin{equation}
 v=\Exp_a\left(F\PT_{c\to a}\Log_c(b)\right),\qquad
 y_i=\Exp_{x_i}\left(B_i\diag(m)B_i^*\Log_{x_i}(v)\right).
 \label{eq:de}
\end{equation}
Here $B_i$ is an orthonormal real tangent frame and $B_i^*$ takes metric coordinates. The binomial mask has Bernoulli rate $CR$, with one uniformly selected coordinate forced active. Thus its marginal activation probability is $CR+(1-CR)/D$, rather than $CR$, and distinct mask bits are dependent~\cite{zaharie2009,novak2026}. Full crossover returns the mutant exactly. The generation is frozen during proposal construction; each trial replaces its target if its loss is no larger. We fix $F=0.7$ and study $CR\in\{0.2,0.8,1\}$.

For a SOMA migrant with frozen origin $x$ and leader $L$, candidates are
\begin{equation}
 z_j=\Exp_x\left(t_jB_x\Pi_j B_x^*\Log_x(L)\right),\qquad
 x^+\in\arg\min_{y\in\{x,z_1,\ldots,z_J\}}f(y).
 \label{eq:soma}
\end{equation}
The leader and population are frozen during a migration, and all path points use the same origin~\cite{novak2026}. Fresh Bernoulli masks with activation $PRT\in\{0.1,0.3,1\}$ are sampled for every point. All-zero masks are permitted and their duplicate evaluations consume budget. The main path is $t_j=0.2j$, $j=1,\ldots,10$, and both algorithms use $N=\max(12,4D)$ individuals. These are matched experimental settings, not the benchmark profile of Ref.~\cite{novak2026}.

We compare four standard geometric constructions (Table~\ref{tab:constructions}). The geodesic variant uses exact geometric operations; the other three replace one or more of them with computationally cheaper alternatives. On a torus, the ambient variant means raw angular subtraction with a wrapped endpoint. Partial DE crossover remains the geodesic tangent-mask operation in Eq.~\eqref{eq:de} in every variant. The ambient partial-crossover variant is therefore a controlled mutation-geometry baseline, not a reproduction of every conventional box-based DE implementation.

\begin{table}[!htbp]
\centering\small
\caption{The four geometric constructions. SOMA moves from a migrant toward its leader and needs no donor-to-base transport, so the projected-transport variant applies to DE only. The crossover mask is identical across variants.}
\label{tab:constructions}
\begin{tabular}{llll}
\toprule
Variant & Difference & Carry to mutation base (DE) & Endpoint map\\
\midrule
Geodesic & Logarithm $\Log_c(b)$ & Parallel transport & Exponential map\\
Retraction & Logarithm $\Log_c(b)$ & Parallel transport & Retraction\\
Projected transport & Logarithm $\Log_c(b)$ & Projection onto $T_a$ & Exponential map\\
Ambient & Phase-aligned chord $b-c$ & Projection onto $T_a$ & Retraction\\
\bottomrule
\end{tabular}
\end{table}

All curved variants use full masks and matched nominal masks ($CR=0.8$ or $PRT=0.3$); additional sparse rates are tested in the geodesic variant. Frames are obtained by transporting an initial reference frame along selected radial geodesics. Their role is to define which local directions a mask can remove. This explicit convention is neither a globally canonical frame nor an eigenvector crossover rule. Some manifolds admit global frames, but a path-independent parallel frame is not available on a curved region in general. Coincident constructions are omitted on flat domains. No success-history adaptation, learned covariance basis, inverse-QFI mutation scaling, local solver, or newly named optimizer is introduced.

Table~\ref{tab:settings} collects the common experimental settings.

\begin{table}[!htbp]
\centering\small
\caption{Common experimental settings.}
\label{tab:settings}
\begin{tabular}{ll}
\toprule
Setting & Value\\
\midrule
Population size & $N=\max(12,4D)$\\
DE & DE/rand/1/bin, $F=0.7$, $CR\in\{0.2,0.8,1\}$\\
SOMA & All-to-one, $PRT\in\{0.1,0.3,1\}$, path $t_j=0.2j$, $j=1,\ldots,10$\\
Budget and checkpoints & $5000D$ evaluations; incumbent also recorded at $1000D$\\
Instances and seeds & 20 instance blocks $\times$ 10 seeds per configuration and operator setting\\
Runs & 84,800 main and 1,200 focused SOMA runs (86,000 in total)\\
\bottomrule
\end{tabular}
\end{table}

Eight focused configurations additionally test a path extended to $3$, reuse of one mask along that path, and the profile of Ref.~\cite{novak2026}, $N=10D$, $PRT=0.1$, $t_j=0.11j$, $j=1,\ldots,27$. These focused comparisons use five instance blocks and ten seeds. Changing to the paper profile changes population and path settings jointly, so it is a profile comparison rather than a single-factor ablation.

\subsection{How geometry changes the three operations}
\label{sec:props}

Before the measurements, it helps to state what each geometric change should do to the search. The three mechanisms below are elementary, but they predict where an algorithm that is unaware of the geometry wastes function evaluations (FEs). We label them M1--M3 and refer to them in Section~\ref{sec:results}. Each was verified numerically; the checks are part of the reproducibility package~\cite{zenodo2026}.

\paragraph{M1: A difference must mean the same thing everywhere in the population}
In $\Reals^D$, the DE difference $b-c$ is the same vector wherever it is applied, and over random donor pairs its second moment is exactly twice the population covariance (Eq.~\eqref{eq:cov}). This is what makes DE self-scaling: a spread population proposes long steps, and a contracted one proposes short steps. On a periodic angle, raw subtraction measures the long way round whenever two donors sit on opposite sides of the numerical $0/2\pi$ seam. This can happen even in a compact population: from $359^\circ$ to $1^\circ$, raw subtraction gives $-358^\circ$, whereas the shortest angular difference is $+2^\circ$. When the population lies in a coordinate interval shorter than half a period that does not cross the seam, raw and shortest-path differences coincide and the Euclidean law holds exactly. Otherwise some raw differences are almost a full turn, the algorithm keeps proposing long jumps after it has already found the right basin, and selection rejects most of them, which costs FEs without progress. A second, related effect appears in widely dispersed populations: even shortest-path differences are then no longer differences of coordinates measured from one reference point, so the covariance law no longer describes them (Section~\ref{sec:results}). On a curved domain such as a sphere, a difference read at $c$ is not tangent at $a$ and must be transported. The error of ignoring this grows smoothly, roughly with the curvature times the squared population radius~\cite{pennec2006}. Consistent geometry should therefore matter little for compact populations and most when the population is spread across a seam or a strongly curved region, which is exactly when raw arithmetic can stop the population from contracting.

\paragraph{M2: A mask acts on parameters, not on what the parameters do}
Binomial crossover and the SOMA perturbation mask switch coordinates off, which always shortens the step in parameter space. For a quantum circuit, however, the relevant movement is the change of the quantum state, measured by the pullback metric $g^\Phi$ (Eq.~\eqref{eq:pullback}). If every parameter moves the state independently, as for product-state angles, where $g^\Phi$ is diagonal, masking can only reduce the local rate of state change. If two parameters partly cancel each other, as in the circuit of Fig.~\ref{fig:overview}(b), switching one of them off removes the cancellation, and a shorter parameter step moves the state further. Such local amplification is possible only when $g^\Phi$ has nonzero off-diagonal entries; it does not occur for every proposal. For finite steps the nonlinear circuit map can increase endpoint distances even when $g^\Phi$ is diagonal, which we measure separately. On layered circuits an angular mask is therefore not the conservative operation it appears to be, and an algorithm that measures its step size only in parameters misjudges how far it actually moves. Masks applied in an orthonormal frame of the state space avoid the effect by construction.

\paragraph{M3: Exact full-mask SOMA can copy its leader}
In Euclidean space the SOMA candidate $z=x+t\Pi(L-x)$ moves along a straight line toward the leader, and with a full mask the path point at $t=1$ is the leader itself~\cite{novak2026}. The exact geodesic version keeps this property, because $\Exp_x(\Log_x L)=L$. If no point on a migrant's path beats the leader, selection replaces the migrant by an exact copy of the leader. A population of identical copies can no longer move, since every difference is zero, so the remaining FEs are spent re-evaluating the same point. Sparse masks ($PRT<1$) and damping retractions reduce the probability of exact leader copying and thereby the risk of collapse. This explains why full-mask SOMA can collapse, not how often it does; the frequency is measured in Section~\ref{sec:results}.

\subsection{Geometry-only mechanisms and quantum diagnostics}

For a fixed Euclidean population with sample covariance $C$ (denominator $N-1$), uniformly sampled distinct ordered donors satisfy
\begin{equation}
 \E[(x_r-x_s)(x_r-x_s)^\top\mid P]=2C.
 \label{eq:cov}
\end{equation}
To test its manifold interpretation, we draw 24 points around a reference $\mu$, express $\Log_\mu(x)$ in a reference frame, and compare $2C$ with the second moment of all 552 ordered vectors $\PT_{c\to\mu}\Log_c(b)$. This is a geometric reference panel over all population pairs, not the exact conditional mutant distribution after DE's target and base exclusions. We record relative Frobenius error, covariance error, and mean drift separately. Local radii are $0.02,0.1,0.4,0.8,1.2$, with five independently sampled populations per radius. A sixth, reference-distribution condition draws points directly from each domain's declared sampler. Seven domain settings give 210 populations. The reference distributions differ across domains and do not constitute a matched curvature experiment.

For SOMA, the Euclidean prediction at mask activation $p$ is
\begin{equation}
 \frac{\E\|z-L\|^2}{\|x-L\|^2}=1-p(2t-t^2).
 \label{eq:contraction}
\end{equation}
We compare the actual leader-centered logarithms with a flat tangent proxy using exactly the same masks, at $p\in\{0.1,0.3,1\}$ and $t\in\{0.5,1,1.5,2.5\}$. The actual-minus-proxy residual isolates geometric distortion from finite mask-sampling error. Small-triangle holonomy tests additionally check the connection: the sphere control approaches sectional curvature one, while the chosen holomorphic projective plane approaches four under Eq.~\eqref{eq:fs}.

Equation~\eqref{eq:contraction} is quadratic in path position $t$, despite the linear Euclidean proposal map. Curvature changes finite geodesic triangles, so a masked path need not follow that flat distance prediction. In contrast, the unmasked exact construction satisfies $\Exp_x(\Log_x L)=L$: a $t=1$ step reaches the leader and does not overshoot it. Beyond the shortest-path region, wrapping can further alter distance. The contraction residual is measured directly instead of being inferred solely from curvature sign.

Quantum diagnostics compare a full proposal tangent $u$ with its masked version $u_m$. For angular masks, the relevant local quantities are $u^\top g^\Phi u$ and $u^\top M g^\Phi M u$, where $M=\diag(m)$. Although $\|Mu\|\leq\|u\|$, no general ordering holds between these two state-space quantities when $g^\Phi$ has off-diagonal terms. In an orthonormal intrinsic frame, by contrast, $\|B M B^*u\|_g\leq\|u\|_g$. A diagonal positive-semidefinite pullback also precludes this particular local amplification. Anisotropy alone is therefore insufficient to cause it.

For seed zero of each instance, up to 16 candidate evaluation positions are sampled uniformly using an independent diagnostic random stream. Central differences, with phase alignment of neighboring rays, estimate pushed tangent norms; actual endpoint changes use exact Fubini--Study distances. Amplification requires an absolute norm increase greater than $10^{-7}$. For input-dependent quantum regression circuits, squared distances and norms are averaged over the fixed input states before taking the square root. Diagnostics do not affect proposals or selection. Selection flags retain algorithm-specific meanings: DE target acceptance, or SOMA selection as the final improving best path point.

\paragraph{Selection and computational cost}
We analyze the recorded end-to-end run time at the common final evaluation budget. Paired exact/alternative runtime ratios are averaged on a log scale within instance blocks; pointwise 95\% bootstrap intervals resample those blocks. These are descriptive runtime comparisons from the same cluster execution. Whole-run replacements per 1,000 trial evaluations mean weakly improving target replacements for DE, including ties, and strictly improving migrant replacements after path selection for SOMA. The different semantics are retained, so the rates are compared within an operator. In the stored quantum proposal sample we separately count strict parent improvements, $f(y)<f(x)-10^{-12}$, and selected candidates. This avoids interpreting accepted duplicates as useful progress. Equal-wall-time comparisons are reconstructed from these timings and the stored best-so-far traces (Section~\ref{sec:design}).

\subsection{Problems, budgets, and analysis}
\label{sec:design}

Table~\ref{tab:design} summarizes the 33 configurations; \ref{app:objectives} specifies the objectives. Generic quantum circuits use exact statevectors, with no shots or device noise. Unrestricted state problems are small geometric controls, not efficient replacements for general VQE. Restricted VQE losses are normalized against Hamiltonian spectral bounds; a zero lower bound is not assumed attainable by the ansatz. QAOA uses $1-\langle C\rangle/C_{\max}$, where $C_{\max}$ is the graph's classical MaxCut value. Quantum regression reports training mean squared error only.

\begin{table}[!htbp]
\centering\small
\caption{Study configurations. $D$ is the real search dimension; paired angular and intrinsic representations have the same feasible state family and initial physical states. Counts total 33. In figures, HEA denotes hardware-efficient-ansatz VQE and PQC the parameterized-quantum-circuit regression model.}
\label{tab:design}
\begin{tabular}{p{39mm}p{49mm}p{35mm}r}
\toprule
Family & Search domain / representation & Sizes and dimensions & Count\\
\midrule
Rotated quadratic & $\Reals^D$ & $D=6,12$ & 2\\
Angular synchronization & Flat torus & $D=6,12$ & 2\\
Rayleigh quotient & Sphere; projective space & $S^6,S^{12}$; $\CP^3,\CP^7,\CP^{15}$ & 5\\
Target-matrix distance & Affine-invariant SPD & Orders $3,5$; $D=6,15$ & 2\\
Full-state VQE & $\CP^{2^n-1}$ & $n=2,3,4$; $D=6,14,30$ & 3\\
Product-state VQE & $(\CP^1)^n$ and polar angles & $n=4,8$; $D=8,16$ & 4\\
Fixed-CZ VQE & Unitary image of product states; angles & $n=4,8$; $D=8,16$ & 4\\
Single-edge QAOA & $\CP^1$ and two angles & Two qubits; $D=2$ & 2\\
Layered QAOA & Parameter torus & $(n,p)=(6,2),(8,3),(10,3)$; $D=4,6,6$ & 3\\
Hardware-efficient VQE & Parameter torus, two layers & $n=4,6$; $D=16,24$ & 2\\
Fidelity regression & $\CP^{2^n-1}$, 32 inputs & $n=2,3$; $D=6,14$ & 2\\
Circuit regression & Parameter torus, 16 inputs & $n=3,4$; $D=12,16$ & 2\\
\bottomrule
\end{tabular}
\end{table}

The main experiment uses 20 randomized instance blocks and ten optimization seeds per configuration and operator setting, giving 84,800 runs. The single-edge QAOA objective itself is fixed; its 20 blocks randomize initialization rather than graph instances. Other instance distributions are defined by the problem generators in the reproducibility package. Identical seeds and population sizes give identical initial populations across operator ablations, with matched physical states across equivalent representations. Proposal random streams are shared within an algorithm; matching random draws does not match the realized trial points or selection paths.

Each run continues to $5000D$ candidate evaluations and also records the incumbent at $1000D$. These are paired checkpoints from one trajectory, not independent runs. Initialization, duplicate proposals, and invalid geometry attempts consume budget. With 1,200 focused SOMA runs, the total is 86,000 trajectories and 172,000 requested endpoint observations.

Primary contrasts compare ambient and geodesic variants separately for each problem, operator, budget, and matched mask level. For each instance, we average the ten paired seed differences, then use an exact two-sided sign-flip test over the 20 instance-mean differences, following nonparametric practice for comparing evolutionary algorithms~\cite{derrac2011}. The test assumes sign exchangeability under its null. We apply Holm correction~\cite{holm1979} to the 248 primary tests jointly, including both budgets. Separate exploratory Holm families comprise 264 mask, 204 primitive, 100 representation, and 48 focused-profile comparisons. Confidence intervals are 95\% percentile bootstrap intervals from 20,000 resamples of instance blocks; they are pointwise, not simultaneous intervals. Focused five-block tests have a minimum attainable two-sided $p$-value of $0.0625$ and are treated descriptively.

Raw losses are retained. Inferential comparisons and loss ratios use $\max(f,10^{-10})$ to avoid tests of numerical residuals near zero. This precision convention and exact, rather than Monte Carlo, sign enumeration were chosen during postprocessing; the experimental manifest and primary contrast structure were fixed before execution. We label a mean difference as materially nonzero for summary counts when its magnitude is at least $10^{-6}$; this is a reporting threshold, not a physical accuracy target. Precision-floor sensitivity is reported in \ref{app:validation}. We compare losses within a common problem and budget and do not pool raw losses into an overall optimizer ranking.

\paragraph{Fixed-target, wall-time and diversity views}
Final losses at two budgets can hide qualitative differences, so three further views of the same records are reported. None requires new runs. First, for the 20 configurations with certified minima, a run succeeds when its best-so-far loss reaches $10^{-6}$. The expected running time~\cite{auger2005,hansen2021coco} is $\mathrm{ERT}=\sum_r e_r/n_{\rm succ}$, where $e_r$ is the first recorded hit or the full budget for unsuccessful runs; it is reported in units of $D$. Second, each primary geometric contrast is repeated at a reconstructed equal wall time; no run was terminated on a time limit. For a paired run, the per-evaluation times $\tau_{\rm amb}$ and $\tau_{\rm geo}$ are the recorded wall time divided by evaluations. Both variants receive the wall time that the faster variant needs for $bD$ evaluations, and each best-so-far trace is read at its last checkpoint inside that time. An optional constant $c$ is added to both per-evaluation costs as a proxy for an expensive objective; $c\to\infty$ recovers evaluation matching. Traces are recorded on a logarithmic grid, so this reading is conservative for the variant whose matched count falls between checkpoints. The same sign-flip test and a fresh Holm family are used for each value of $c$. Third, the final population radius, the mean intrinsic distance of members from the incumbent, is compared between geometric variants.

\section{Results}
\label{sec:results}

All 86,000 planned trajectories completed, providing 172,000 requested checkpoint endpoints with no missing tasks or duplicate successes. The geometric, masking, and selection comparisons below use this complete set. Numerical geometry penalties are retained in the budgets and reported with the affected SPD variant. The detailed integrity and resource audit is given in \ref{app:validation}.

\subsection{Classical population statistics hold locally and fail globally}

\begin{figure}[!htbp]
\centering\includegraphics[width=.85\textwidth]{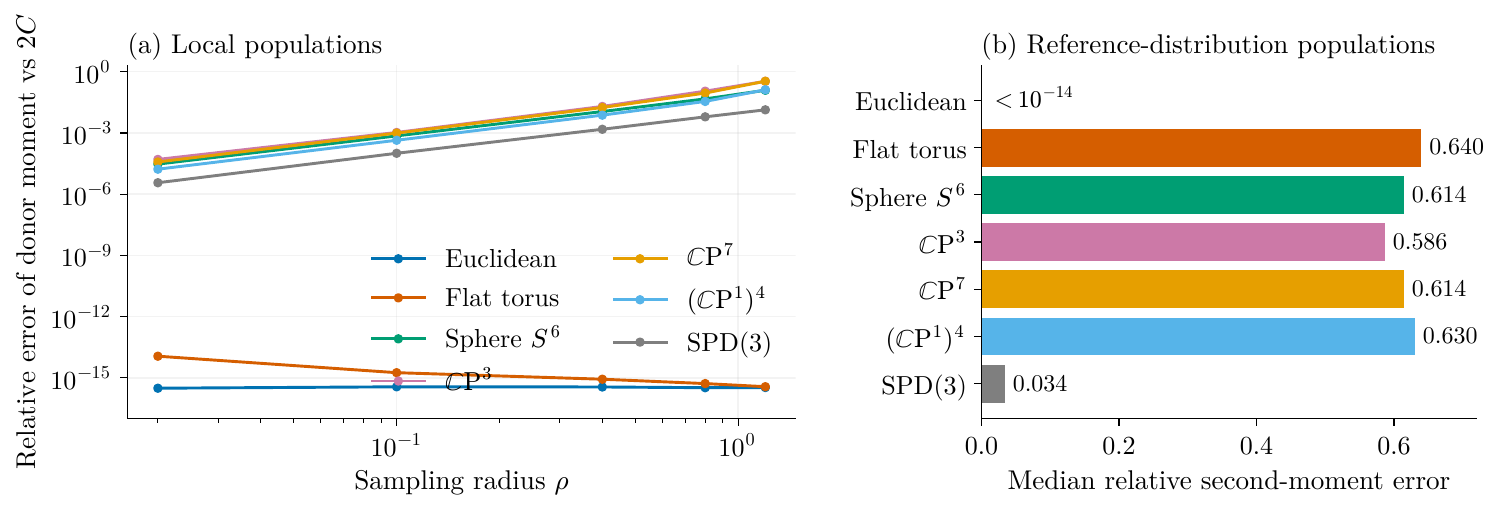}
\caption{Transported donor second moments relative to the Euclidean $2C$ law. Local points have tangent radii sampled between $0.4\rho$ and $\rho$. Curves and bars show medians over five populations; all 552 distinct ordered donor pairs are enumerated within each population. The torus is locally flat yet has a large discrepancy for its dispersed uniform reference population.}
\label{fig:donors}
\end{figure}

Figure~\ref{fig:donors} shows recovery of Eq.~\eqref{eq:cov} for local populations. Euclidean relative errors remain near machine precision. Local torus populations also recover the law, as their pairwise differences stay within a consistent shortest-path region. Curved domains show radius-dependent deviations. Log--log fits over radii $0.02,0.1,0.4$ give slopes between 1.99 and 2.04 for the sphere, the two projective controls, the product projective control, and SPD. These descriptive slopes support a second-order local discrepancy; they are not a convergence theorem for the algorithms.

The reference-distribution condition behaves differently. Median relative second-moment errors are 0.640 on the flat torus, 0.614 on the sphere, 0.586 and 0.614 on $\CP^3$ and $\CP^7$, 0.630 on $(\CP^1)^4$, and 0.0336 on the sampled SPD control. The torus discrepancy is particularly informative: it occurs without curvature or nontrivial parallel transport. Pairwise shortest logarithms cannot generally be written as differences of two fixed-reference logarithms for a dispersed wrapped population. Thus global angular covariance and local donor covariance answer different questions. These domain-specific reference samplers do not support ranking the magnitude of curvature from the listed errors.

Curved transported donor directions also acquire nonzero mean at the reference point. Median normalized mean drift in the reference-distribution condition is 0.0554 for $\CP^3$, 0.0525 for $\CP^7$, and 0.107 for the sphere; Euclidean and torus controls remain at numerical zero. Small-triangle transport independently recovers curvature proxies 1.0001 on the sphere and 4.0011 on the selected holomorphic projective planes at radius 0.02 (\ref{app:validation}).

\begin{figure}[!htbp]
\centering\includegraphics[width=.85\textwidth]{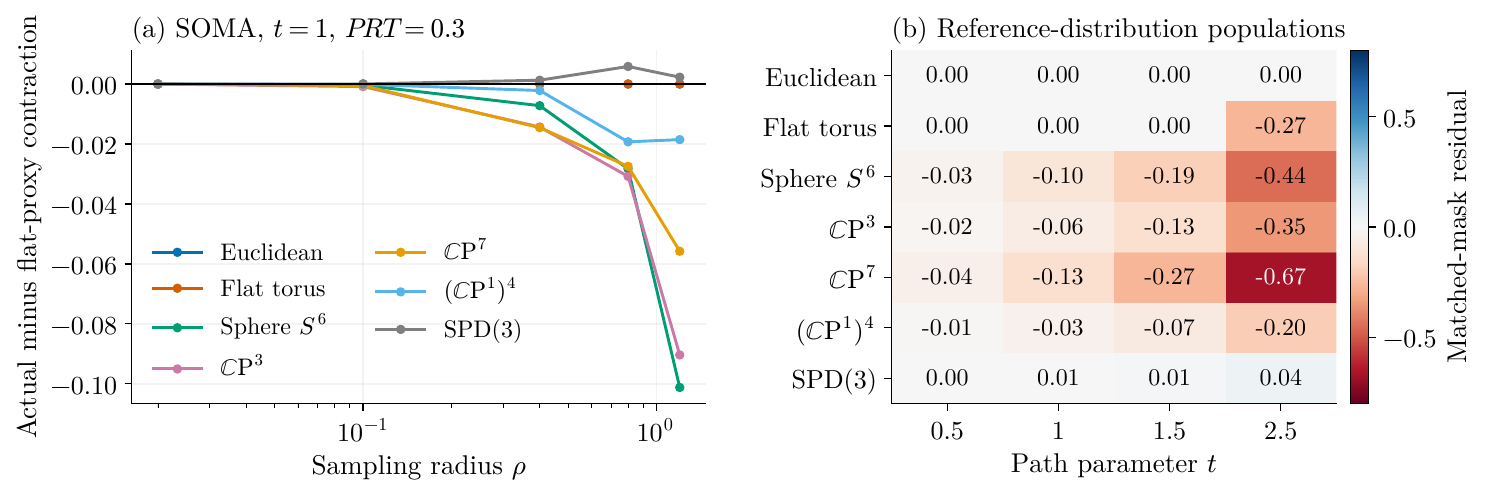}
\caption{Actual-minus-flat-proxy normalized squared leader distance for SOMA masks with $PRT=0.3$. The same masks are used in both quantities. (a) Median residual at $t=1$ versus population radius. (b) Median residual versus path position for reference-distribution populations. Curvature, quotient wrapping, and finite overshoot change the Euclidean contraction prediction; the residual contains no fitness selection.}
\label{fig:soma}
\end{figure}

The masked SOMA residual also separates local and finite-scale effects (Fig.~\ref{fig:soma}). At $t=1$, $PRT=0.3$, and local radius 0.02, median matched-mask residuals have magnitude below $2.2\times10^{-5}$. In the reference-distribution condition they are $-0.0612$ for $\CP^3$, $-0.1307$ for $\CP^7$, $-0.0986$ on the sphere, and $+0.00623$ on SPD. A negative residual means smaller leader distance than the matched flat proxy predicts; it does not by itself imply lower objective value. The torus residual remains zero at $t=1$ but departs under sufficiently long overshoot, illustrating wrapping without curvature.

\subsection{Shortest-path DE improves 16 of 31 configurations at full crossover}

\begin{figure}[p]
\centering\includegraphics[width=.97\textwidth,height=.76\textheight,keepaspectratio]{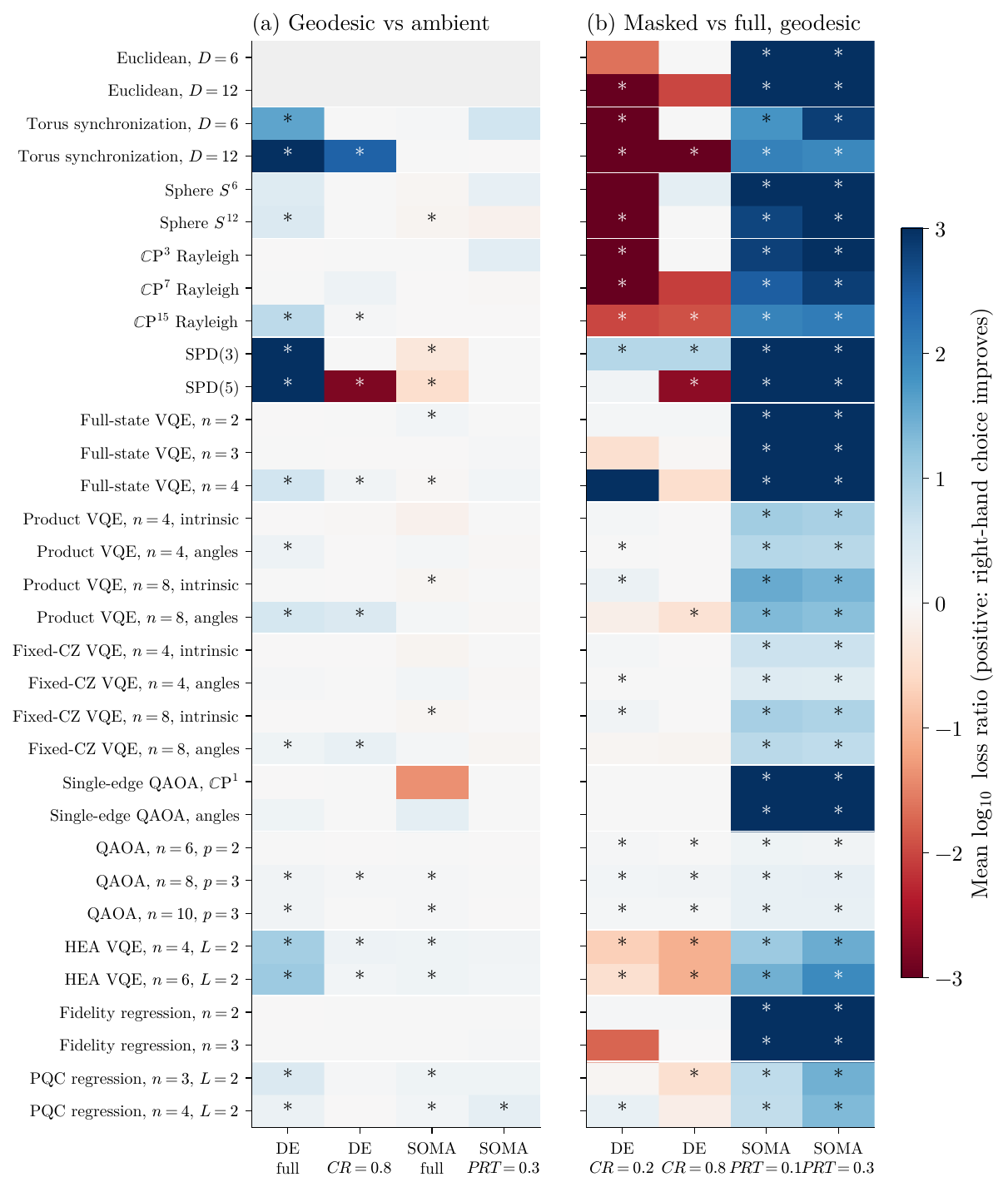}
\caption{Geometry and mask interactions across all 33 configurations at $5000D$. Colors give the instance-balanced mean log$_{10}$ loss ratio after a $10^{-10}$ precision floor: positive values favor geodesic over ambient in (a), and masked over full updates in (b). Color saturates at $\pm3$ decades. An asterisk denotes a raw-scale mean loss difference of magnitude at least $10^{-6}$ with Holm-adjusted $p<0.05$, using separate primary and mask test families over both budgets. Color and significance summarize different effect scales. Gray cells are omitted redundant Euclidean variants.}
\label{fig:map}
\end{figure}

Figure~\ref{fig:map} summarizes all configurations at $5000D$. For full-crossover DE, geodesic mutation yields a Holm-significant, materially lower mean loss than ambient mutation in 16 of the 31 non-Euclidean configurations, with no significant reversal in this primary comparison. This count describes the selected operators and budget, not a universal superiority result. At $CR=0.8$, the corresponding counts are eight improvements and one reversal. Full-mask SOMA has seven improvements and six reversals; at $PRT=0.3$ only one primary geometry contrast meets both criteria. Full-mask SOMA is a structural control with different diversity dynamics from its masked versions.

On six-qubit hardware-efficient VQE, full-crossover DE mean normalized loss decreases from 0.09449 with raw angular donor differences to 0.009186 with shortest-path differences. The paired mean gain is 0.08530, with pointwise 95\% interval $[0.07721,0.09360]$ and Holm-adjusted $p=0.000473$. For ten-qubit depth-three QAOA, the corresponding losses are 0.12852 and 0.10500, giving gain 0.02351, interval $[0.01984,0.02709]$. These improvements concern the flat parameter torus, so they cannot be attributed to implementing the curved state-space metric.

The intrinsic projective controls also show finite-budget improvements. For $\CP^{15}$ Rayleigh minimization, full-crossover DE decreases mean loss from $4.82\times10^{-4}$ to $7.10\times10^{-5}$, a gain of $4.11\times10^{-4}$ with interval $[3.35,4.98]\times10^{-4}$. Full-state four-qubit VQE decreases from $9.23\times10^{-4}$ to $2.80\times10^{-4}$. These are unrestricted state controls with certified spectral minima.

The mask results rule out a general prescription to set $CR=1$ for quantum problems (Table~\ref{tab:losses}). Full crossover is clearly best on the $\CP^{15}$ Rayleigh problem and on six-qubit hardware-efficient VQE, whereas partial crossover performs better on layered QAOA and circuit regression. Preserving the complete geometric direction is therefore not universally beneficial; it must be balanced against the changes in search diversity and accepted moves produced by partial masks.

\begin{table}[!htbp]
\centering\small\setlength{\tabcolsep}{3pt}
\caption{Mean final losses at $5000D$ for selected geodesic variants (200 runs per cell). DE columns give $CR$ and SOMA columns give $PRT$. Values are descriptive raw means; values below $10^{-10}$ are shown at the analysis precision floor. VQE and generic QAOA losses are spectral-bound-normalized objectives, not certified restricted-ansatz gaps.}
\label{tab:losses}
\fittable{\begin{tabular}{lrrrrrrr}
\toprule
Problem & $D$ & DE 1 & DE .2 & DE .8 & SOMA 1 & SOMA .1 & SOMA .3 \\
\midrule
Euclidean & 12 & $<10^{-10}$ & $0.00188$ & $1.71\!\times\!10^{-8}$ & $0.438$ & $3.49\!\times\!10^{-4}$ & $1.42\!\times\!10^{-5}$ \\
$\CP^{15}$ Rayleigh & 30 & $7.10\!\times\!10^{-5}$ & $0.00535$ & $0.00381$ & $0.161$ & $0.00325$ & $0.00287$ \\
Full-state VQE (4q) & 30 & $2.80\!\times\!10^{-4}$ & $3.43\!\times\!10^{-4}$ & $5.08\!\times\!10^{-4}$ & $0.193$ & $5.21\!\times\!10^{-4}$ & $2.98\!\times\!10^{-4}$ \\
Product VQE (8q), intrinsic & 16 & $0.00773$ & $0.00381$ & $0.00773$ & $0.167$ & $0.0121$ & $0.0159$ \\
Product VQE (8q), angles & 16 & $0.00435$ & $0.00516$ & $0.00884$ & $0.126$ & $0.0128$ & $0.0157$ \\
QAOA (10q, p=3) & 6 & $0.105$ & $0.0909$ & $0.0913$ & $0.166$ & $0.1$ & $0.101$ \\
HEA VQE (6q) & 24 & $0.00919$ & $0.0221$ & $0.0795$ & $0.125$ & $0.00523$ & $0.00382$ \\
PQC regression (4q) & 16 & $0.00526$ & $0.00249$ & $0.00651$ & $0.00884$ & $0.00162$ & $6.85\!\times\!10^{-4}$ \\
$\mathrm{SPD}(5)$ & 15 & $8.79\!\times\!10^{-9}$ & $<10^{-10}$ & $0.00177$ & $1.11$ & $<10^{-10}$ & $<10^{-10}$ \\
\bottomrule
\end{tabular}
}
\end{table}

Full-mask SOMA is consistently poor on several otherwise smooth controls (Table~\ref{tab:losses}). Its grid includes $t=1$, which copies the leader exactly under Exp. If no later point improves on that leader, selection can eliminate migrant diversity. At $5000D$, final population radius is at most $10^{-8}$ in 90.5\% of ten-qubit QAOA full-mask runs, compared with 0\% at $PRT=0.3$. The corresponding rates are 34.5\% versus 0\% for six-qubit hardware-efficient VQE, 10.5\% versus 0\% for four-qubit full-state VQE, and 23\% versus 0\% for the 12-dimensional quadratic. Collapse is not universal and is not specific to curvature. On QAOA, mean loss decreases from 0.16626 with full masks to 0.10089 at $PRT=0.3$. The mechanism is consistent with the leader-copying structure, but the study does not include a matched grid excluding only $t=1$, so that single causal contribution is not isolated.

\begin{figure}[!htbp]
\centering\includegraphics[width=.9\textwidth]{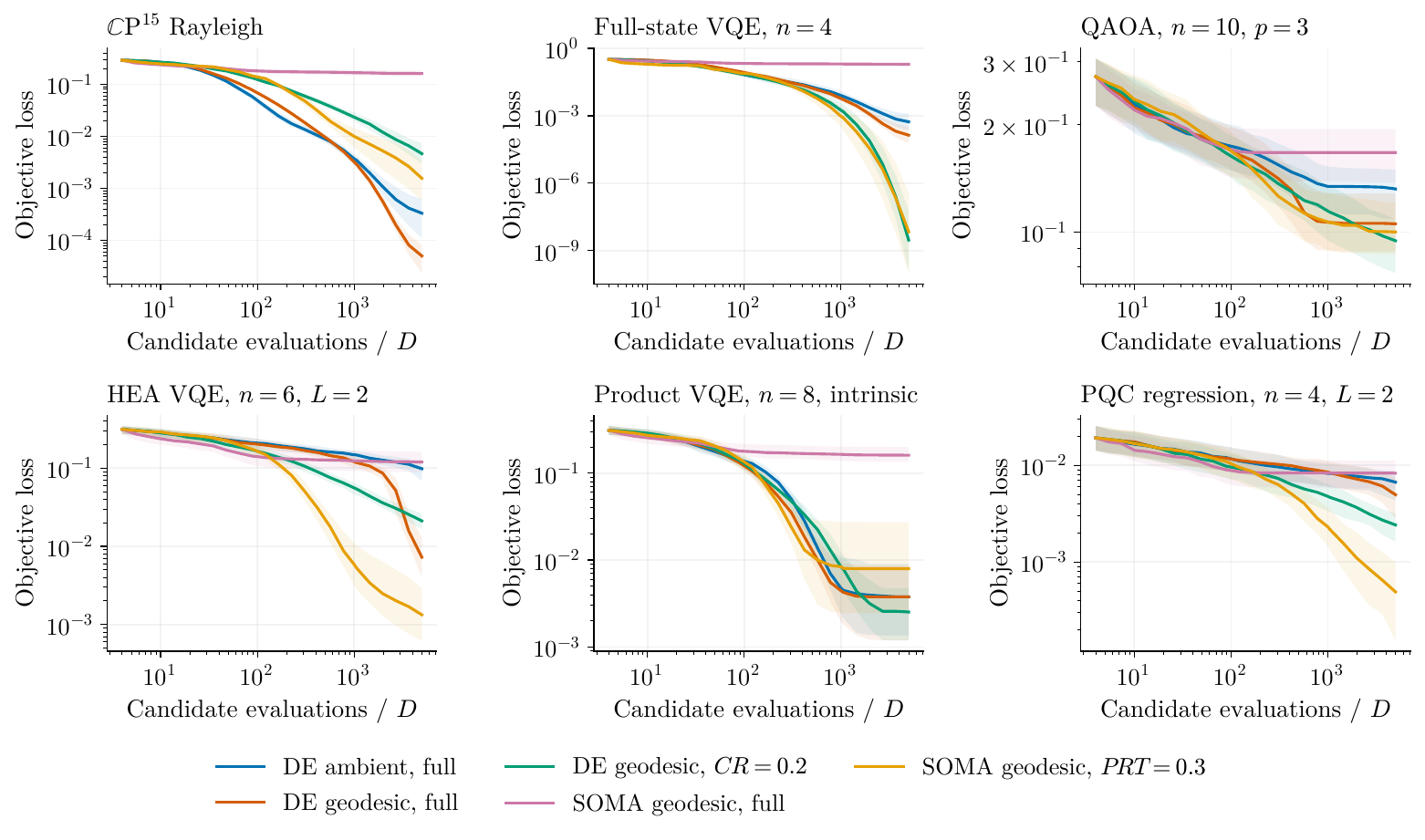}
\caption{Representative best-so-far trajectories under fixed operators. Lines show medians and shading shows interquartile ranges across 200 runs, rather than confidence intervals. Losses are floored at $10^{-10}$ for plotting. Both checkpoints belong to the same run. Full masks suit some DE objectives but not others; full-mask SOMA often stagnates. The highlighted problems were selected for interpretation, not as an independent confirmatory panel.}
\label{fig:trajectories}
\end{figure}

\subsection{Computational cost, success probability and population diversity}

\begin{figure}[!htbp]
\centering\includegraphics[width=\textwidth]{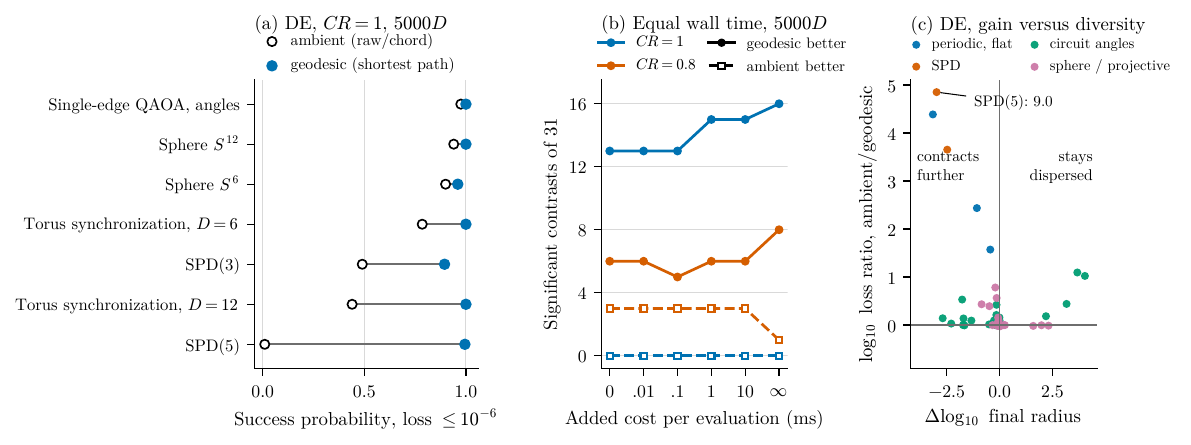}
\caption{Additional views of the same 86,000 runs. (a) Probability of reaching loss $\le10^{-6}$ within $5000D$ evaluations for full-crossover DE, shown for the certified-minimum configurations where the two variants differ by more than 0.02; the other 11 certified-minimum configurations with both variants have equal or both-zero success. (b) Number of Holm-significant, materially nonzero primary contrasts at equal wall time, as a function of a constant cost added to every objective evaluation; $\infty$ is evaluation matching. Solid lines count geodesic improvements and dashed lines count ambient improvements. (c) Mean $\log_{10}$ loss ratio (ambient/geodesic, positive favors geodesic) against the change in $\log_{10}$ final population radius (geodesic minus ambient) for every DE geometric contrast; the SPD(5) point at 9.0 decades is drawn at the top edge.}
\label{fig:outcomes}
\end{figure}

Exact geometry is not free. In the intrinsic curved variants, geometric operations take 52--87\% of the recorded runtime (median over runs), against 4--12\% for the layered circuits, where objective evaluation dominates. With these exact statevector objectives one evaluation costs only 0.01--1.0\,ms, so the overhead matters. Figure~\ref{fig:outcomes}(b) therefore repeats the primary contrasts at a reconstructed equal wall time, assuming a constant cost per evaluation within each run. For full-crossover DE, geodesic geometry remains better in 13 of 31 configurations with the recorded costs and in 15 of 31 once each evaluation costs 1\,ms or more, with no significant reversal at any cost. At $CR=0.8$ the count falls from 8 improvements and 1 reversal at matched evaluations to 6 and 3 at matched time; the additional reversals are the cheap-objective cases where a retraction or chord step is both faster and adequate. Full-mask SOMA remains balanced (7 improvements, 6--7 reversals) and $PRT=0.3$ SOMA shows almost no geometric effect under either matching. When an objective evaluation costs much more than the geometric overhead, as is typical for shot-based hardware estimates, the evaluation-matched comparison becomes the relevant one. With cheap simulators the wall-time view applies, and it reduces but does not remove the full-crossover advantage.

\begin{table}[!htbp]
\centering\small\setlength{\tabcolsep}{4pt}
\caption{Fixed-target view for full-crossover DE at $5000D$: success probability for loss $\le10^{-6}$ and, in parentheses, expected running time $\mathrm{ERT}/D$ (total evaluations per success, in units of $D$). 200 runs per cell; ``--'' marks variants not defined on flat domains. Representative configurations with certified minima are shown; all 20 are in the reproducibility package~\cite{zenodo2026}.}
\label{tab:success}
\fittable{\begin{tabular}{lcccc}
\toprule
Problem & Ambient & Geodesic & Retraction & Proj.\ transport\\
\midrule
Torus synchronization, $D=6$ & 0.79 (2,254) & 1.00 (571) & -- & --\\
Torus synchronization, $D=12$ & 0.44 (9,642) & 1.00 (1,936) & -- & --\\
Sphere $S^6$ & 0.90 (1,050) & 0.96 (707) & 0.94 (759) & 0.92 (948)\\
Sphere $S^{12}$ & 0.94 (2,537) & 1.00 (1,648) & 0.99 (1,617) & 0.98 (1,939)\\
SPD(3) & 0.49 (6,035) & 0.90 (1,389) & 0.91 (1,231) & 0.01 (998,667)\\
SPD(5) & 0.01 (499,334) & 0.99 (3,603) & 0.99 (2,599) & 0.69 (6,152)\\
$\mathbb{C}\mathrm{P}^7$ Rayleigh & 1.00 (1,582) & 1.00 (1,574) & 1.00 (1,525) & 1.00 (1,532)\\
Full-state VQE, $n=3$ & 1.00 (1,882) & 1.00 (1,854) & 1.00 (1,755) & 1.00 (1,803)\\
Fidelity regression, $n=3$ & 1.00 (1,504) & 1.00 (1,369) & 1.00 (1,304) & 1.00 (1,304)\\
Single-edge QAOA, angles & 0.97 (337) & 1.00 (197) & -- & --\\
\bottomrule
\end{tabular}
}
\end{table}

Final mean losses understate some of the largest effects (Fig.~\ref{fig:outcomes}(a), Table~\ref{tab:success}). On 12-dimensional torus synchronization, success rises from 0.44 with raw angular differences to 1.00 with shortest-path differences, and ERT falls from 9,642$D$ to 1,936$D$. On SPD(5) it rises from 0.01 to 0.99, an ERT ratio of about 140. Sphere controls gain a few percentage points. The projective quantum controls that both variants solve ($\CP^7$, three-qubit full-state VQE, fidelity regression) have equal success and ERT within 10\%: the primitive hardly matters once every run converges. Retraction often has the smallest ERT among the curved variants, which agrees with its step damping (Section~\ref{sec:retraction}). Projected transport fails on SPD(3), with success 0.01, consistent with the non-isometric step scale discussed there.

The diversity view shows two contrasting patterns associated with the same gain (Fig.~\ref{fig:outcomes}(c)). On torus synchronization and SPD, geodesic DE at full crossover ends with a population radius up to three decades \emph{smaller} than the ambient variant (12-dimensional torus, $-3.2$ decades; SPD(5), $-3.0$). On the torus, raw differences that cross the angular seam keep proposing long, mostly useless steps, so the ambient population cannot contract onto the optimum. This is the failure of the donor-moment law described by M1 (Section~\ref{sec:props}). On SPD the ambient chord has no seam problem; its step scale differs from the affine-invariant one, and the population again fails to contract, but the data do not isolate which property is responsible. On the layered circuits the final radius is instead \emph{larger} under shortest-path differences, by 3.2--4.0 decades on average in log scale for both hardware-efficient VQE sizes and three-qubit circuit regression. In six-qubit hardware-efficient VQE, for example, the median final radius is 4.9 against 0.089, and 23.5\% of ambient runs end fully collapsed (radius below $10^{-8}$) against none with shortest-path differences. Here the ambient population collapses prematurely and keeps a high weak-acceptance rate from near-ties, which offers an explanation for the observation in Section~\ref{sec:selection} that lower loss can accompany a lower acceptance rate. Across all DE contrasts, gain and radius change are moderately anticorrelated (Spearman $-0.44$), but no single diversity direction accounts for the benefit. These are associations between final radius and final loss; the runs do not isolate diversity as the cause.

\subsection{Approximate geometry can outperform exact updates at partial crossover}
\label{sec:retraction}

Isolating the geometric primitive changes the interpretation (Table~\ref{tab:primitives}). On $\CP^{15}$ with full-crossover DE, exact geodesic mutation outperforms the tested retraction and projected-transport alternatives. At $CR=0.8$, the ordering reverses: the retraction and projected-transport means are approximately 0.00268 and 0.00272, compared with 0.00381 for the exact construction. These effects are Holm-significant in the exploratory primitive family. The outcome is not explained by tangent-space validity alone: endpoint maps and non-isometric transport alter finite-step scale and interact with crossover and selection.

For normalized sphere and projective retractions, $R_p(u)=(p+u)/\|p+u\|$ traverses an angle $\arctan\|u\|$ along the same initial direction, rather than $\|u\|$. Thus a retraction can damp a large step; at $t=1$ it need not reach the SOMA leader. On $\SPD(5)$, full-mask geodesic SOMA has mean loss 1.113, whereas the retraction variant has mean 0.434. The masked versions of both reach numerical precision on this distance objective. This does not make retraction more geometrically exact; it shows that exact interpolation and finite-budget optimization quality are different properties.

\begin{table}[!htbp]
\centering\small\setlength{\tabcolsep}{3pt}
\caption{Primitive ablations at $5000D$. ``Exact'' uses Log, parallel transport where needed, and Exp. The same mask and selection rule are retained. Means use 200 runs; $p_H$ is Holm-adjusted over 204 primitive contrasts, including both budgets. Projected transport applies only to DE.}
\label{tab:primitives}
\fittable{\begin{tabular}{lrrrrrr}
\toprule
Problem & Operator & Mask & Alternative & Exact & Alternative & $p_H$ \\
\midrule
$\CP^{15}$ & DE & 1 & Retraction & $7.10\!\times\!10^{-5}$ & $1.20\!\times\!10^{-4}$ & $0.0007$ \\
$\CP^{15}$ & DE & .8 & Retraction & $0.00381$ & $0.00268$ & $0.0004$ \\
$\CP^{15}$ & DE & 1 & Projection transport & $7.10\!\times\!10^{-5}$ & $1.91\!\times\!10^{-4}$ & $0.0004$ \\
$\CP^{15}$ & DE & .8 & Projection transport & $0.00381$ & $0.00272$ & $0.0004$ \\
$\mathrm{SPD}(5)$ & SOMA & 1 & Retraction & $1.11$ & $0.434$ & $0.0004$ \\
\bottomrule
\end{tabular}
}
\end{table}

The SPD projected-transport variant also exposes a numerical limitation. Reusing a donor tangent matrix at another SPD base is a valid ambient tangent identification, but it is not an isometry of the affine-invariant metric. The resulting very large metric-scaled steps can cause finite-precision matrix exponential/logarithm calculations to lose positive definiteness. All 29,830 invalid attempts occur in this variant: 24 for order three and 29,806 for order five. The order-five masked variant contributes 28,814 attempts, or approximately 0.192\% of that variant's budget. No failed proposal is silently repaired by eigenvalue clipping, and all penalties remain in the reported outcomes.

\subsection{Lower loss does not require a higher acceptance rate}
\label{sec:selection}

\begin{figure}[!htbp]
\centering\includegraphics[width=\textwidth]{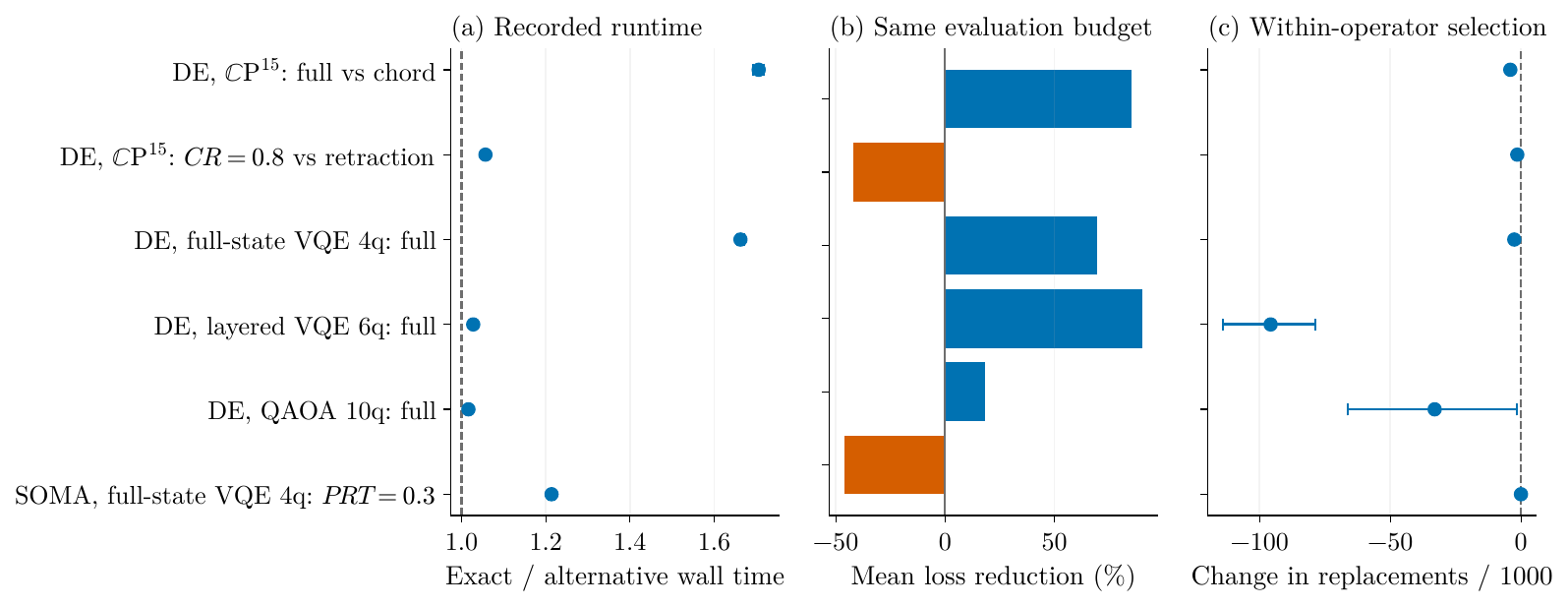}
\caption{Measured cost, quality, and selection at $5000D$. Every row compares exact geometry with one declared alternative, retaining operator, mask rate, and initialization. (a) Instance-balanced geometric mean runtime ratios with pointwise 95\% bootstrap intervals. (b) Raw mean loss reduction, $100(1-\bar f_{\rm exact}/\bar f_{\rm alternative})$, at equal candidate budget; negative values favor the alternative. (c) Difference in whole-run replacements per 1,000 trials, with block-bootstrap intervals. DE includes equal-loss acceptance; SOMA counts one improving migrant replacement per path. Only within-row comparisons have matched selection semantics.}
\label{fig:cost}
\end{figure}

The measured cost of exact geometry depends strongly on the problem representation (Fig.~\ref{fig:cost}). On $\CP^{15}$ Rayleigh minimization at $CR=1$, the paired exact/chord runtime ratio is 1.706, with pointwise 95\% interval $[1.693,1.719]$. Mean loss decreases by 85.3\%, from $4.82\times10^{-4}$ to $7.10\times10^{-5}$, yet replacements decrease from 64.4 to 60.4 per 1,000 trials. Exact transport and movement buy lower final loss at this evaluation budget, not a larger accepted fraction.

The periodic correction is less expensive in the tested layered circuits. For six-qubit hardware-efficient VQE, shortest-path full-crossover DE costs a factor 1.028 in recorded runtime, interval $[1.022,1.034]$, and reduces mean normalized loss by 90.3\%. Its weak replacement rate falls from 127.2 to 31.4 per 1,000 trials. The sampled strict improvement rates also fall, from 4.06\% to 2.50\% (320 proposals per variant). These observations show that counting surviving candidates alone would give the wrong impression of the quality gain. Weak acceptance can include many near-stagnation ties, and even a strict improvement count omits the sizes of improvements and the states reached.

The comparison also identifies a case where exact geometry is unfavorable in both observed cost and final loss. At $CR=0.8$ on $\CP^{15}$, the exact construction costs 1.057 times the retraction variant, interval $[1.052,1.061]$, while its mean loss is 0.00381 rather than 0.00268. Thus the retraction is both faster and more accurate at this common evaluation budget; this ordering reverses at full crossover.

For four-qubit full-state VQE with SOMA at $PRT=0.3$, exact movement costs 1.214 times the chord/retraction alternative, interval $[1.208,1.220]$. The sampled selected-point rates are 12.81\% and 10.94\%, but the primary final-loss comparison has no corrected evidence of an improvement. Its means are $2.98\times10^{-4}$ and $2.04\times10^{-4}$, respectively. Higher sampled survival therefore does not establish a quality benefit either. Table~\ref{tab:cost} retains both operators' selection definitions and reports strict proposal improvements separately. These are final-budget cost--quality observations; the wall-time-matched view is given in Fig.~\ref{fig:outcomes}(b).

\begin{table}[!htbp]
\centering\small\setlength{\tabcolsep}{3pt}
\caption{Selection and runtime details. Each paired entry is exact/alternative. Wall time is the median over 200 runs; replacement rate is the mean per 1,000 trial evaluations. The mask column means $CR$ for DE and $PRT$ for SOMA. Strict parent-improvement percentages use only stored quantum probes (320 per variant here) and tolerance $10^{-12}$; ``--'' denotes no quantum proposal probes for the textbook objective. A strict parent improvement need not be the SOMA path winner.}
\label{tab:cost}
\fittable{\begin{tabular}{lrrrrrr}
\toprule
Problem & Alg. & Mask & Alternative & Wall s & Repl./1000 & Strict (\%) \\
\midrule
CP15 & DE & 1 & Chord/raw & 31.99/18.70 & 60.4/64.4 & -- \\
CP15 & DE & 0.8 & retraction & 60.17/57.05 & 24.5/25.9 & -- \\
Full-state VQE, 4q & DE & 1 & Chord/raw & 33.73/20.25 & 58.8/61.3 & 4.69/6.88 \\
Layered VQE, 6q & DE & 1 & Chord/raw & 80.42/78.27 & 31.4/127.2 & 2.50/4.06 \\
QAOA, 10q & DE & 1 & Chord/raw & 34.78/34.26 & 372.6/405.6 & 9.38/8.12 \\
Full-state VQE, 4q & SOMA & 0.3 & Chord/raw & 18.43/15.21 & 100.0/100.0 & 65.94/64.69 \\
\bottomrule
\end{tabular}
}
\end{table}

\subsection{Coordinate masks can increase quantum-state displacement}

\begin{figure}[!htbp]
\centering\includegraphics[width=.9\textwidth]{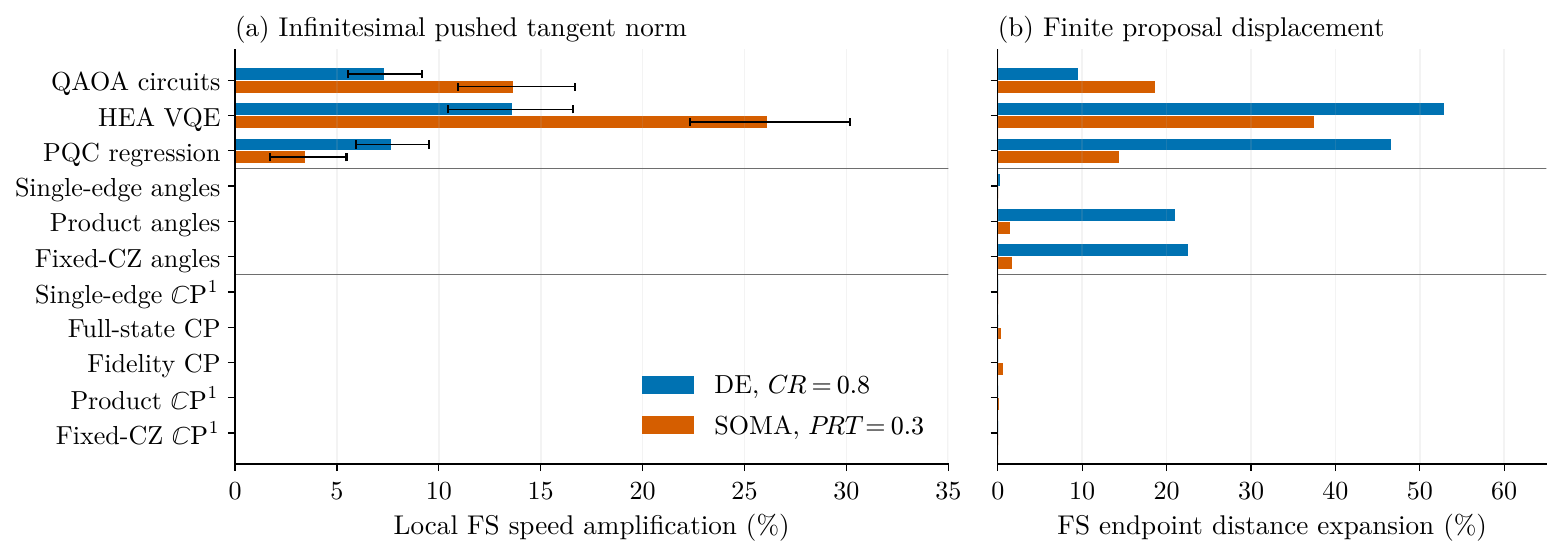}
\caption{Masking and quantum-state displacement for the main geodesic variants at $CR=0.8$ and $PRT=0.3$. (a) Fraction of sampled proposals whose masked pushed tangent norm exceeds the unmasked norm by more than $10^{-7}$; bars have pointwise 95\% instance-block bootstrap intervals. (b) Fraction whose actual FS endpoint distance increases under the mask. Probes are pooled with equal counts across the tested sizes within a family and include unselected proposals and realized zero/full masks. Local and finite-step effects are distinct.}
\label{fig:motion}
\end{figure}

We distinguish the initial rate of state change in the proposed direction, measured by the local pushed tangent norm, from the total state displacement produced by a finite update, measured by the FS endpoint distance. The former reveals coupling between parameter directions, while the latter also includes nonlinear and periodic effects along the update. Figure~\ref{fig:motion} compares the matched geodesic rates $CR=0.8$ and $PRT=0.3$. For generic QAOA, angular masking increases local pushed FS norm in 70/960 DE probes (7.29\%; pointwise block-bootstrap interval $[5.52,9.17]\%$) and 131/960 SOMA probes (13.65\%; $[10.94,16.67]\%$). In hardware-efficient VQE the rates are 87/640 (13.59\%; $[10.47,16.56]\%$) and 167/640 (26.09\%; $[22.34,30.16]\%$). Circuit regression gives 49/640 (7.66\%) and 22/640 (3.44\%). In all cases, the masks still contract the declared parameter-tangent norm.

No local amplification is observed in the intrinsic projective or product-projective controls. It is also absent in the angular product-state and single-edge QAOA controls, whose analytic pullback metrics are diagonal. The product metric is
\begin{equation}
 g^\Phi=\tfrac14\bigoplus_{j=1}^n\diag(1,\sin^2\theta_j).
 \label{eq:productmetric}
\end{equation}
A fixed entangling unitary preserves this metric, so it does not generate local mask amplification by itself. The empirical zero rates in both product and fixed-CZ angular controls agree with this identity. Entanglement alone is therefore not a sufficient explanation for off-diagonal pullback coupling.

The pullback probes also distinguish possible rank degeneracy from what was observed. At the absolute eigenvalue threshold $10^{-7}$, each hardware-efficient VQE size has 17 numerically rank-deficient probes out of 200, with minimum ranks 14/16 and 22/24. Most recorded points have full numerical rank. Rank deficiency of this metric is a known signature of redundant circuit parameters~\cite{haug2021,larocca2023}. The circuit optimizers use flat parameter geometry and never invert the pullback matrix, so these observations do not constitute a benchmark of intrinsic optimization with a singular metric. The rank-one circuit in Fig.~\ref{fig:overview}(b) supplies an exact explanatory example of redundancy.

Finite endpoint displacement is a separate observable. At the same matched rates, DE masking increases actual FS endpoint distance in 52.81\% of hardware-efficient VQE probes and 46.56\% of circuit-regression probes, well above their local speed-amplification rates. Even the diagonal-metric product-angle controls have finite-distance expansion in 20.94\% of DE probes. Local contraction does not imply global contraction after a nonlinear periodic circuit map, especially for long directions. Intrinsic SOMA also has a few finite-distance reversals during overshoot despite no local norm amplification.

Selection does not turn amplification into a sufficient criterion for usefulness. Among the matched QAOA probes, three of 137 DE-accepted candidates and ten of 53 SOMA path-selected candidates exhibit local amplification. Conversely, many amplifying proposals are rejected. These small conditional counts, and the different selection definitions, do not support causal claims that amplification helps or harms optimization. They show that the geometric event is not eliminated completely by selection.

\subsection{Crossover changes which representation is preferable}

The representation controls share objective functions, real dimension, population size, and initial physical states. Yet intrinsic and angular optimizers do not have equivalent proposal distributions. On eight-qubit product-state VQE, geodesic DE at $CR=0.2$ improves mean loss from 0.005157 in angles to 0.003810 intrinsically, with gain 0.001346, pointwise interval $[0.001056,0.001684]$, and representation-family adjusted $p=0.000191$. With full crossover the direction reverses descriptively: angular DE has mean 0.004351 and intrinsic DE 0.007731, although the corrected $p=0.0545$ does not cross 0.05. For the fixed-CZ image, full-crossover angular DE is significantly better in the corresponding exploratory family, whereas $CR=0.2$ favors the intrinsic representation (Table~\ref{tab:representations}).

\begin{table}[!htbp]
\centering\small
\caption{Matched feasible-set comparisons at $5000D$ for eight-qubit models. Positive gain is angular minus intrinsic mean loss. Initial physical states are paired. $p_H$ is adjusted over 100 representation contrasts across both budgets.}
\label{tab:representations}
\fittable{\begin{tabular}{lrrrrr}
\toprule
Ansatz (8q) & Operator & Intrinsic & Angles & Gain & $p_H$ \\
\midrule
Product & DE, CR=1 & $0.00773$ & $0.00435$ & $-0.00338$ & $0.0545$ \\
Product & DE, CR=.2 & $0.00381$ & $0.00516$ & $0.00135$ & $0.0002$ \\
Product & SOMA, PRT=.3 & $0.0159$ & $0.0157$ & $-2.54\!\times\!10^{-4}$ & $1.0000$ \\
Fixed CZ & DE, CR=1 & $0.0196$ & $0.0147$ & $-0.00487$ & $0.0188$ \\
Fixed CZ & DE, CR=.2 & $0.0136$ & $0.0156$ & $0.00195$ & $0.0002$ \\
Fixed CZ & SOMA, PRT=.3 & $0.0363$ & $0.0323$ & $-0.00408$ & $1.0000$ \\
\bottomrule
\end{tabular}
}
\end{table}

The focused SOMA panel gives additional descriptive evidence, not a definitive profile recommendation (\ref{app:focused}). On $\CP^{15}$, reusing one mask over the extended path increases mean loss from 0.002835 to 0.007417. The profile of Ref.~\cite{novak2026} has mean 0.01545 versus 0.003246 for its matched-instance main $PRT=0.1$ baseline. Other objectives favor that profile. These comparisons combine finite path sampling, effective search dimension, and migration count; five instance blocks cannot resolve small corrected $p$-values. They identify follow-up questions rather than establish a universally better SOMA setting.

\FloatBarrier
\section{Discussion}

\subsection{What transfers from Euclidean DE and SOMA theory}

The Euclidean moment and masking identities remain useful reference laws, but their scope must be stated. Pairwise shortest logarithms recover affine differences in small neighborhoods. Once the population spans multiple quotient branches, that representation can fail even on a flat torus. Curvature adds base-dependent transport, local moment distortion, and mean drift. These are separate mechanisms (M1, Section~\ref{sec:props}): the torus's large global moment error cannot be used as evidence of curvature, and projective holonomy cannot be reproduced by merely wrapping angles. The fixed-target results show that the global torus failure is not a curiosity of the objective-free panel. It is the regime in which raw angular DE fails to contract and loses most of its success probability.

For population covariance, the practical implication is to declare the reference point and transport convention. A covariance of reference logarithms describes one tangent representation of a population. It is not automatically the covariance of the difference vectors DE actually samples. The experiments support this local distinction without introducing a covariance estimator or a new adaptive rule.

For SOMA, Eq.~\eqref{eq:contraction} predicts local geometric behavior before selection. Its curvature and wrapping residuals do not predict the sign of an energy improvement. Moreover, the sampling grid and the leader-copying event (M3) can dominate diversity dynamics~\cite{pluhacek2021}. The full-mask results should therefore be read as a useful structural control, not as evidence against the SOMA family or as an evaluation of its adaptive variants.

\subsection{Implications for VQA optimization}

The most immediately supported modification in the tested circuit settings is consistent shortest-path subtraction using valid coordinate periods. It is inexpensive, adding about 2--5\% to recorded runtime in the layered circuits, and it can substantially improve full-crossover DE, partly by preventing premature collapse of the population. Exact intrinsic updates are a different trade. With cheap simulated objectives their overhead is large enough to remove some of their advantage at equal wall time. When an evaluation is much more expensive than the geometric operations, as for shot-based estimates on quantum hardware, the overhead becomes small and evaluation-matched comparisons apply. The benefit is not universal at every mask rate, and it does not implement quantum-state natural geometry. The circuit pullback provides the further distinction: eliminating a coordinate can remove cancellation between coupled parameter directions and increase the physical state change.

The practical crossover conclusions are problem specific: full crossover is best on the projective Rayleigh problem, partial crossover on layered QAOA. Neither curvature nor observed mask amplification is sufficient to prescribe full crossover. The original Euclidean operator--selection distinction~\cite{novak2026} remains central: geometry describes what is proposed, while selection, diversity, and objective structure determine its value at a finite budget.

Nor does a more intrinsic domain guarantee better optimization. Exact updates remove geometric inconsistencies relative to a chosen metric, but the metric does not encode the objective's descent structure. Retractions and non-isometric transport can change effective step scale in ways that help some combinations of objective and mask. Product-state and angular representations can reach the same states while exploring them differently. These outcomes argue for matched feasible-set and matched-mask ablations before interpreting an apparent improvement as a consequence of curvature handling.

The fixed-unitary control further separates state entanglement from metric coupling. A fixed entangler changes the energy restricted to a product-state family while preserving its induced FS metric. Its optimization behavior can therefore change without a change in intrinsic geometry. A claim that entanglement necessarily requires a dense metric, or necessarily requires $CR=1$, would be too broad for these data.

\subsection{Scope and future work}

The study isolates mechanisms by holding the DE and SOMA operators fixed. Which complete optimizer performs best on variational quantum problems is a different question, addressed by dedicated benchmarks of classical, adaptive, and evolutionary methods~\cite{lavrijsen2020,bonetmonroig2023,novak2026noisy,illesova2025vha,illesova2025bench,bezdek2025,novak2026cec,novak2025reliable}. The present results indicate how such optimizers can be improved: geometry-consistent differences, intrinsic masks, and the diagnostics developed here can be combined with SHADE-family DE~\cite{tanabe2013,tanabe2014,piotrowski2026}, adaptive SOMA~\cite{kadavy2021,skanderova2023}, CMA-ES~\cite{hansen2001}, or quantum natural gradient~\cite{stokes2020}.

Sampling noise is a particularly well-motivated next setting. On quantum hardware, every objective value is a finite-shot estimate, so selection itself becomes stochastic, which acts directly on the near-tie replacements, the leader copying of M3, and population collapse studied here. Population methods are comparatively robust in noisy variational landscapes~\cite{novak2026noisy,novak2025reliable}, and how geometry and masking interact with noisy selection is an open question. Larger instances, where gradients concentrate~\cite{mcclean2018} and the pullback metric can become rank deficient~\cite{haug2021,larocca2023}, and coordinate-wise methods that exploit the trigonometric structure of circuits~\cite{nakanishi2020,ostaszewski2021}, which are subject to the same mask effects as M2, are equally relevant.

The mechanisms also suggest targeted experiments: removing only the leader-copying grid point in full-mask SOMA, comparing exact and retraction updates at matched step lengths, and probing controlled tangent directions at common circuit states to relate mask amplification to the pullback metric. Learned or eigenvector-aligned frames for intrinsic masks, and timing with realistic objective costs, would further refine where exact geometry pays off.

\section{Conclusions}

The benchmark supports four operational conclusions. First, shortest-path differences are a low-cost correction that should be the default for angular parameters. They reduce full-crossover six-qubit VQE loss by 90.3\% and raise 12-dimensional torus synchronization success from 44\% to 100\%. Second, exact projective updates can improve loss substantially, but their cost and their interaction with crossover must be measured: the projective retraction wins at $CR=0.8$ while exact movement wins at $CR=1$, and part of the advantage disappears at equal wall time on cheap objectives. Third, the gains are associated with two contrasting diversity patterns, so acceptance rates and final radii alone cannot certify a better operator. Fourth, masks can locally increase quantum-state motion only when the pullback metric couples parameters (M2), and finite steps can increase it more often. Geometry, masking, and selection are therefore separate benchmark factors, each with directly observable consequences.

\section*{Data availability}
The code, processed data, raw run records, and figure sources needed to reproduce all results are openly available on Zenodo~\cite{zenodo2026} at \url{https://doi.org/10.5281/zenodo.24516254}.
\section*{Funding}
{\sloppy This project has received funding from the Research Council of Lithuania (LMTLT), agreement No.\ \mbox{S-ITP-25-7}. This work was supported by the European Union under the REFRESH -- Research Excellence For REgion Sustainability and High-tech Industries project, project number CZ.10.03.01/\allowbreak 00/22\_003/0000048, via the Operational Programme Just Transition. This research was also supported by research grant SGS No.\ SP2026/063 of VSB--Technical University of Ostrava, Czech Republic. This work was supported by the Ministry of Education, Youth and Sports of the Czech Republic through the e-INFRA CZ (ID:90254).\par}

\section*{Acknowledgments}
Computational resources were provided by IT4Innovations National Supercomputing Center through project FTA-26-86.

\section*{Declaration of competing interest}
The authors declare that they have no known competing financial interests or personal relationships that could have appeared to influence the work reported in this paper.

\appendix
\section{Geometric formulas and validation}
\label{app:geometry}
\paragraph{Curvature of the test domains} The unit sphere has sectional curvature one. Under Eq.~\eqref{eq:fs}, projective holomorphic planes have curvature four, while real sectional curvatures in $\CP^m$, $m\geq2$, range from one to four. Affine-invariant SPD geometry has nonpositive curvature, including flat commuting directions; it is not a constant strictly negative-curvature space. Holonomy is used to validate the connection, not as a separately isolated cause of optimizer performance.

For projective states, let $\widehat q=q\exp[-i\arg(p^\dagger q)]$, $w=(I-pp^\dagger)\widehat q$, and $\alpha=\operatorname{atan2}(\|w\|,|p^\dagger q|)$. The implemented primitives are
\begin{equation}
 \Log_p(q)=\alpha\frac{w}{\|w\|},\qquad
 \Exp_p(u)=p\cos\|u\|+u\frac{\sin\|u\|}{\|u\|},\qquad
 R_p(u)=\frac{p+u}{\|p+u\|}.
\end{equation}
The continuous zero-vector limits are used. With $h=\Log_p(q)/\alpha$, parallel transport expressed at $\widehat q$ is
\begin{equation}
 \widehat u=u+\big[(\cos\alpha-1)h-\sin\alpha\,p\big](h^\dagger u),
\end{equation}
followed by the phase conversion to the given representative $q$. The Hermitian coefficient $h^\dagger u$ remains complex; retaining only its real part would fail horizontal transport. At orthogonal rays, the shortest logarithm is not unique. The code uses a deterministic representative branch and counts such encounters. Sphere antipodes and torus half-period displacements are handled analogously. Beyond-injectivity tangent inputs are permitted and recorded, not clipped.

On $\SPD(k)$, with symmetric tangents $U,V$, the metric and maps are
\begin{align}
 g_P(U,V)&=\tr(P^{-1}UP^{-1}V),\\
 \Log_P(Q)&=P^{1/2}\log(P^{-1/2}QP^{-1/2})P^{1/2},\\
 \Exp_P(U)&=P^{1/2}\exp(P^{-1/2}UP^{-1/2})P^{1/2},\\
 R_P(U)&=P+U+\tfrac12UP^{-1}U.
\end{align}
Exact parallel transport is $U\mapsto EUE^\top$, with $E=P^{1/2}(P^{-1/2}QP^{-1/2})^{1/2}P^{-1/2}$. Projected transport on SPD retains the ambient symmetric matrix; this differs from the metric isometry. No eigenvalue floor is used to repair failed numerical proposals.

\subsection{Validation and numerical conventions}
\label{app:validation}

\begin{figure}[!htbp]
\centering\includegraphics[width=.85\textwidth]{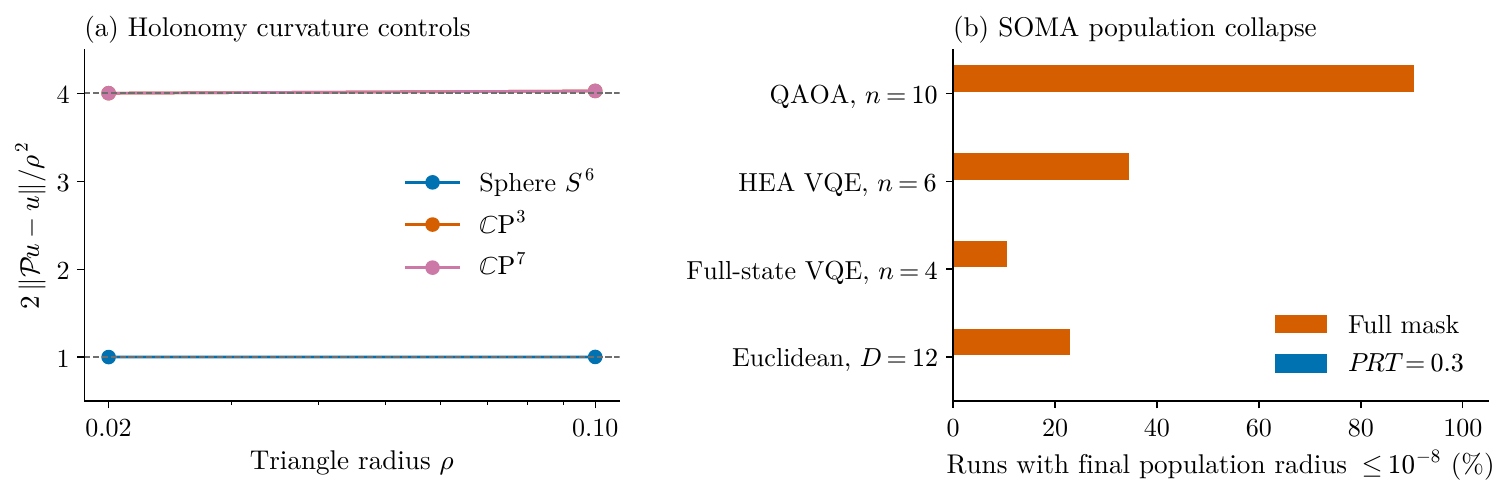}
\caption{Additional mechanism checks. (a) Small-triangle holonomy approaches the known sphere and holomorphic-projective curvatures; the normalization uses approximate area $\rho^2/2$. (b) Fraction of selected SOMA runs with final population radius at most $10^{-8}$. Collapse rates are descriptive and do not isolate the effect of the $t=1$ grid point from the other full-mask dynamics.}
\label{fig:holonomy}
\end{figure}

The geometric primitives were verified by unit tests covering Log--Exp recovery, metric-preserving and inverse transport, projective phase invariance, exclusion of the vertical phase direction, known-curvature holonomy, cut-locus handling, the local covariance law, equivalent quantum representations, and pullback metrics (Fig.~\ref{fig:holonomy}(a)). All 86,000 runs completed with their full evaluation budgets. Re-evaluating 3,440 stratified final points with the objective implementation reproduces the recorded losses to $1.8\times10^{-14}$, and paired angular and intrinsic initial populations have identical objective values to $8.3\times10^{-16}$. The only failed proposals are the 29,830 penalized numerical failures of the SPD projected-transport variant (Section~\ref{sec:retraction}); every final point is finite and feasible to $4.4\times10^{-16}$.

Across both budgets, 87 primary contrasts meet the Holm threshold and the $10^{-6}$ material-effect convention with a $10^{-10}$ precision floor. A $10^{-12}$ floor leaves this classification unchanged, and a $10^{-8}$ floor changes one classification (86 remain). Values below the reporting threshold are not interpreted as physical accuracy on hardware. The study used 13.9 node-hours of worker time on 36-core nodes. Geometry accounted for 53.5\% and objective evaluation for 33.2\% of the aggregated per-run time, so equal evaluation budgets do not imply equal computational cost; these small-simulation timings do not estimate quantum-hardware cost.

\section{Objective definitions and feasible quantum models}
\label{app:objectives}

\paragraph{Textbook controls}
The Euclidean objective is $(x-x_*)^\top A(x-x_*)/100$, with a Gaussian target and randomly rotated eigenvalues geometrically spaced from 1 to 100. Sphere and projective Rayleigh quotients are $x^\dagger Ax$, with a randomly orthogonal or unitary eigenbasis and spectrum $0,\operatorname{linspace}(0.02,1,m-1)$. The torus synchronization objective is a normalized sum of $w_{ij}[1-\cos(e_i-e_j)]$, $e=x-x_*$, on ring and random chord terms, plus $0.2[1-\cos e_0]$ to anchor the common phase. Its certified minimum is zero. The SPD objective is $d(P,P_*)^2/D$, with a randomly oriented target having eigenvalues geometrically spaced from 1 to 100. Initial SPD points are exponentials of scaled Gaussian symmetric matrices. The exact random generators and their deterministic seed keys are in the source snapshot.

\paragraph{Hamiltonian VQE}
The Hamiltonian is
\begin{equation}
 H=\sum_{(i,j)\in E}J_{ij}Z_iZ_j+\sum_i h_iZ_i+\sum_i g_iX_i.
\end{equation}
Edges comprise a ring plus random additional edges; $J_{ij}$ has a random sign and magnitude uniform on $[0.6,1.4]$, $h_i$ is uniform on $[-0.4,0.4]$, and $g_i$ on $[0.4,0.9]$. The loss is $(\langle H\rangle-E_{\min})/(E_{\max}-E_{\min})$. Full-state VQE optimizes a normalized ray. Product VQE uses $\bigotimes_j(\cos(\theta_j/2),e^{i\phi_j}\sin(\theta_j/2))$, represented either by local projective rays or by the redundant angle torus. Fixed-CZ VQE applies a fixed ring of CZ gates to those product states. The unitary preserves their metric and distances, although the restricted energy changes for the same fixed Hamiltonian. Hardware-efficient VQE~\cite{kandala2017} uses two layers of $R_y,R_z$ rotations and fixed ring-CZ entanglers, starting from $|0\rangle^{\otimes n}$.

\paragraph{QAOA}
The unweighted cost is $C=\sum_{(i,j)\in E}(I-Z_iZ_j)/2$. From $|+\rangle^{\otimes n}$, each layer applies $e^{-i\gamma_l C}$ and then $e^{-i\beta_l\sum_iX_i}$. Gamma periods are $2\pi$ divided by the spectral-gap gcd of the integer-valued cost; beta periods are $\pi/2$ in these global-bit-flip-invariant circuits. No arbitrary weighted-cost periodicity is assumed. Generic QAOA remains an angular search.

For the single-edge control, the invariant basis is $|E\rangle=(|00\rangle+|11\rangle)/\sqrt2$ and $|O\rangle=(|01\rangle+|10\rangle)/\sqrt2$. The logical cost is $\diag(0,1)$, the mixer is $2X$, and the reachable ray set is $\CP^1$. We minimize logical even-state probability, with certified minimum zero. In the $(\gamma,\beta)$ coordinates,
\begin{equation}
 g^\Phi=\diag(1/4,4\sin^2\gamma).
\end{equation}
The corresponding angle parameterization has singularities and redundancy, but its initial physical states match the direct projective runs.

\paragraph{Quantum regression}
The fidelity model uses 32 fixed, randomly sampled product-state features after the same fixed CZ embedding. For an unrestricted trainable state $p$, predictions are $|\langle\varphi_a|p\rangle|^2$ and labels are generated by a fixed random target ray. The loss is mean squared error, with attainable minimum zero. This is a small state-geometry training model motivated by overlap-based quantum embeddings~\cite{lloyd2020}, not a reproduction of an application benchmark. The circuit model uses 16 random product inputs, two rotation/CZ layers, and readout $0.6\langle Z^{\otimes n}\rangle+0.4\langle X^{\otimes n}\rangle$. A teacher circuit generates the labels. The mixed readout keeps final $R_z$ parameters observable. There is no test split or claim of generalization.

\section{Complete primary summaries and focused SOMA profiles}
\label{app:focused}

{\footnotesize
\begin{longtable}{lrrrr}
\caption{All primary high-budget mean gains: ambient minus geodesic loss at $5000D$. Positive values favor the geodesic variant. An asterisk indicates Holm-adjusted $p<0.05$ over all 248 primary tests and magnitude at least $10^{-6}$. Values with magnitude below $10^{-6}$ are abbreviated $\approx0$. Full precision, intervals, both budgets, and raw-scale effects are in the reproducibility package~\cite{zenodo2026}. Euclidean variants have no redundant ambient comparison.}\label{tab:allprimary}\\
\toprule Problem & DE 1 & DE .8 & SOMA 1 & SOMA .3 \\
\midrule
\endfirsthead
\caption[]{Primary high-budget mean gains (continued).}\\
\toprule Problem & DE 1 & DE .8 & SOMA 1 & SOMA .3 \\
\midrule
\endhead
\bottomrule
\endfoot
Torus synchronization, D=6 & $0.00171$$^{*}$ & $\approx 0$ & $0.0124$ & $0.00172$ \\
Torus synchronization, D=12 & $0.007$$^{*}$ & $0.00913$$^{*}$ & $0.00827$ & $2.60\!\times\!10^{-4}$ \\
Sphere S6 & $9.99\!\times\!10^{-5}$ & $\approx 0$ & $-0.00252$ & $-6.02\!\times\!10^{-6}$ \\
Sphere S12 & $8.86\!\times\!10^{-5}$$^{*}$ & $\approx 0$ & $-0.0116$$^{*}$ & $-4.18\!\times\!10^{-4}$ \\
$\CP^3$ Rayleigh & $\approx 0$ & $\approx 0$ & $0.00109$ & $\approx 0$ \\
$\CP^7$ Rayleigh & $\approx 0$ & $\approx 0$ & $-0.0018$ & $1.15\!\times\!10^{-6}$ \\
$\CP^1$5 Rayleigh & $4.11\!\times\!10^{-4}$$^{*}$ & $4.44\!\times\!10^{-4}$$^{*}$ & $-0.00596$ & $-2.86\!\times\!10^{-6}$ \\
SPD(3) & $0.102$$^{*}$ & $\approx 0$ & $-0.27$$^{*}$ & $\approx 0$ \\
SPD(5) & $0.2$$^{*}$ & $-0.00177$$^{*}$ & $-0.756$$^{*}$ & $\approx 0$ \\
Full-state VQE, n=2 & $2.36\!\times\!10^{-6}$ & $\approx 0$ & $0.0111$$^{*}$ & $\approx 0$ \\
Full-state VQE, n=3 & $\approx 0$ & $\approx 0$ & $0.00272$ & $-8.43\!\times\!10^{-6}$ \\
Full-state VQE, n=4 & $6.42\!\times\!10^{-4}$$^{*}$ & $1.14\!\times\!10^{-4}$$^{*}$ & $-0.00916$$^{*}$ & $-9.41\!\times\!10^{-5}$ \\
Product VQE, n=4, intrinsic & $1.62\!\times\!10^{-4}$ & $-5.98\!\times\!10^{-4}$ & $-0.0146$ & $6.67\!\times\!10^{-4}$ \\
Product VQE, n=4, angles & $0.00192$$^{*}$ & $-1.79\!\times\!10^{-4}$ & $0.00469$ & $-0.00188$ \\
Product VQE, n=8, intrinsic & $-3.34\!\times\!10^{-5}$ & $-3.16\!\times\!10^{-4}$ & $-0.0199$$^{*}$ & $0.00133$ \\
Product VQE, n=8, angles & $0.0087$$^{*}$ & $0.0132$$^{*}$ & $0.0118$ & $-0.00179$ \\
Fixed-CZ VQE, n=4, intrinsic & $-8.77\!\times\!10^{-4}$ & $-0.00158$ & $-0.0158$ & $0.00248$ \\
Fixed-CZ VQE, n=4, angles & $5.84\!\times\!10^{-4}$ & $-3.62\!\times\!10^{-4}$ & $0.0114$ & $-0.00443$ \\
Fixed-CZ VQE, n=8, intrinsic & $-0.00179$ & $-3.01\!\times\!10^{-4}$ & $-0.0251$$^{*}$ & $-0.0011$ \\
Fixed-CZ VQE, n=8, angles & $0.00463$$^{*}$ & $0.00944$$^{*}$ & $0.0149$ & $-0.00529$ \\
Single-edge QAOA, $\CP^1$ & $\approx 0$ & $\approx 0$ & $5.79\!\times\!10^{-5}$ & $\approx 0$ \\
Single-edge QAOA, angles & $2.87\!\times\!10^{-5}$ & $\approx 0$ & $1.62\!\times\!10^{-4}$ & $\approx 0$ \\
QAOA, n=6, p=2 & $0.0046$ & $-0.00171$ & $0.00584$ & $-0.00356$ \\
QAOA, n=8, p=3 & $0.0228$$^{*}$ & $0.00427$$^{*}$ & $0.0148$$^{*}$ & $5.05\!\times\!10^{-5}$ \\
QAOA, n=10, p=3 & $0.0235$$^{*}$ & $0.00161$ & $0.0126$$^{*}$ & $-7.09\!\times\!10^{-4}$ \\
HEA VQE, n=4, L=2 & $0.0392$$^{*}$ & $0.0119$$^{*}$ & $0.0231$$^{*}$ & $4.24\!\times\!10^{-4}$ \\
HEA VQE, n=6, L=2 & $0.0853$$^{*}$ & $0.0176$$^{*}$ & $0.0381$$^{*}$ & $-6.08\!\times\!10^{-4}$ \\
Fidelity regression, n=2 & $\approx 0$ & $\approx 0$ & $0.00114$ & $\approx 0$ \\
Fidelity regression, n=3 & $\approx 0$ & $\approx 0$ & $1.34\!\times\!10^{-4}$ & $-1.77\!\times\!10^{-5}$ \\
PQC regression, n=3, L=2 & $0.00352$$^{*}$ & $1.64\!\times\!10^{-4}$ & $0.00442$$^{*}$ & $2.90\!\times\!10^{-4}$ \\
PQC regression, n=4, L=2 & $0.00139$$^{*}$ & $-2.09\!\times\!10^{-4}$ & $0.00195$$^{*}$ & $3.99\!\times\!10^{-4}$$^{*}$ \\

\end{longtable}
}
\FloatBarrier

\begin{table}[!htbp]
\centering\small\setlength{\tabcolsep}{3pt}
\caption{Focused SOMA mean losses at $5000D$, using only the same five instance blocks and ten seeds in every cell. Main .3 and .1 refer to the main $PRT$ settings; Path3 uses $PRT=0.3$. The paper profile uses $PRT=0.1$ as in Main .1 but changes population and path grid jointly. Exact five-block sign-flip tests cannot reach $p<0.05$.}
\label{tab:focused}
\fittable{\begin{tabular}{lrrrrr}
\toprule
Problem & Main .3 & Path3 fresh & Path3 reused & Main .1 & Paper profile \\
\midrule
Torus synchronization, D=6 & $0.00284$ & $6.11\!\times\!10^{-4}$ & $0.00451$ & $0.00641$ & $0.00272$ \\
Torus synchronization, D=12 & $0.00764$ & $0.00448$ & $0.00713$ & $0.00643$ & $0.00853$ \\
CP3 Rayleigh & $2.43\!\times\!10^{-7}$ & $2.33\!\times\!10^{-9}$ & $1.20\!\times\!10^{-4}$ & $3.37\!\times\!10^{-4}$ & $1.66\!\times\!10^{-4}$ \\
CP7 Rayleigh & $5.13\!\times\!10^{-4}$ & $3.10\!\times\!10^{-4}$ & $0.00492$ & $8.43\!\times\!10^{-4}$ & $0.00287$ \\
CP15 Rayleigh & $0.00245$ & $0.00284$ & $0.00742$ & $0.00325$ & $0.0155$ \\
Product VQE, n=4, intrinsic & $0.0264$ & $0.0245$ & $0.0251$ & $0.0251$ & $0.0226$ \\
Product VQE, n=8, intrinsic & $0.0135$ & $0.0128$ & $0.0135$ & $0.0129$ & $0.00839$ \\
QAOA, n=6, p=2 & $0.137$ & $0.123$ & $0.122$ & $0.133$ & $0.118$ \\
\bottomrule
\end{tabular}
}
\end{table}

\begin{table}[!htbp]
\centering\small
\caption{Local FS norm amplification counts for selected matched-mask geodesic families. DE uses $CR=0.8$, SOMA uses $PRT=0.3$. Intervals resample the 20 instance blocks and pool tested sizes within a family. Zero observations in the structural controls agree with the metric identity; a degenerate bootstrap interval is not a binomial upper confidence bound. The accompanying CSV contains all families.}
\label{tab:motioncounts}
\begin{tabular}{lrrrrr}
\toprule
Family & Operator & Probes & Amplified & Percent & 95\% CI \\
\midrule
Layered QAOA & DE & 960 & 70 & $7.29$ & $[5.52,9.17]$ \\
Layered QAOA & SOMA & 960 & 131 & $13.65$ & $[10.94,16.67]$ \\
HEA VQE & DE & 640 & 87 & $13.59$ & $[10.47,16.56]$ \\
HEA VQE & SOMA & 640 & 167 & $26.09$ & $[22.34,30.16]$ \\
Circuit regression & DE & 640 & 49 & $7.66$ & $[5.94,9.53]$ \\
Circuit regression & SOMA & 640 & 22 & $3.44$ & $[1.72,5.47]$ \\
Product VQE, angles & DE & 640 & 0 & $0.00$ & $[0.00,0.00]$ \\
Product VQE, angles & SOMA & 640 & 0 & $0.00$ & $[0.00,0.00]$ \\
Fixed-CZ VQE, angles & DE & 640 & 0 & $0.00$ & $[0.00,0.00]$ \\
Fixed-CZ VQE, angles & SOMA & 640 & 0 & $0.00$ & $[0.00,0.00]$ \\
Full-state VQE & DE & 960 & 0 & $0.00$ & $[0.00,0.00]$ \\
Full-state VQE & SOMA & 960 & 0 & $0.00$ & $[0.00,0.00]$ \\
\bottomrule
\end{tabular}

\end{table}

\FloatBarrier
\bibliographystyle{elsarticle-num}
\bibliography{references}

@book{boumal2023,title={An Introduction to Optimization on Smooth Manifolds},author={Boumal, Nicolas},year={2023},publisher={Cambridge University Press},doi={10.1017/9781009166164},url={https://www.nicolasboumal.net/book/}}

@book{absil2008,title={Optimization Algorithms on Matrix Manifolds},author={Absil, P.-A. and Mahony, Robert and Sepulchre, Rodolphe},year={2008},publisher={Princeton University Press}}

@article{storn1997,title={Differential Evolution---A Simple and Efficient Heuristic for Global Optimization over Continuous Spaces},author={Storn, Rainer and Price, Kenneth},journal={Journal of Global Optimization},volume={11},pages={341--359},year={1997},doi={10.1023/A:1008202821328}}

@book{soma2016,title={Self-Organizing Migrating Algorithm: Methodology and Implementation},editor={Davendra, Donald and Zelinka, Ivan},year={2016},publisher={Springer},doi={10.1007/978-3-319-28161-2}}

@article{novak2026,
title = {Linear proposal operators and stochastic search geometry in SOMA and Differential Evolution},
journal = {Swarm and Evolutionary Computation},
volume = {111},
pages = {102595},
year = {2026},
issn = {2210-6502},
doi = {https://doi.org/10.1016/j.swevo.2026.102595},
url = {https://www.sciencedirect.com/science/article/pii/S2210650226003159},
author = {Vojtěch Novák and Ivan Zelinka}
}

@article{opara2018,title={Comparison of mutation strategies in Differential Evolution---A probabilistic perspective},author={Opara, Karol R. and Arabas, Jaros{\l}aw},journal={Swarm and Evolutionary Computation},volume={39},pages={53--69},year={2018},doi={10.1016/j.swevo.2017.12.007}}

@article{opara2019,title={Differential Evolution: A survey of theoretical analyses},author={Opara, Karol R. and Arabas, Jaros{\l}aw},journal={Swarm and Evolutionary Computation},volume={44},pages={546--558},year={2019},doi={10.1016/j.swevo.2018.06.010}}

@article{zaharie2009,title={Influence of crossover on the behavior of Differential Evolution Algorithms},author={Zaharie, Daniela},journal={Applied Soft Computing},volume={9},pages={1126--1138},year={2009},doi={10.1016/j.asoc.2009.02.012}}

@article{colutto2010,title={The CMA-ES on Riemannian Manifolds to Reconstruct Shapes in 3-D Voxel Images},author={Colutto, Sebastian and Fr{\"u}hauf, Florian and Fuchs, Matthias and Scherzer, Otmar},journal={IEEE Transactions on Evolutionary Computation},volume={14},number={2},pages={227--245},year={2010},doi={10.1109/TEVC.2009.2029567}}

@inproceedings{arnold2014,title={On the Use of Evolution Strategies for Optimization on Spherical Manifolds},author={Arnold, Dirk V.},booktitle={PPSN XIII},series={LNCS},volume={8672},pages={882--891},year={2014},url={https://www.cmap.polytechnique.fr/~nikolaus.hansen/proceedings/2014/PPSN/papers/8672/86720882.pdf}}

@article{he2020,title={An evolutionary approach to black-box optimization on matrix manifolds},author={He, Xiaoyu and Zhou, Yuren and Chen, Zefeng and Jiang, Siyu},journal={Applied Soft Computing},volume={97},pages={106773},year={2020},doi={10.1016/j.asoc.2020.106773}}

@misc{lesniewski2025,title={Differential Evolution for Grassmann Manifold Optimization: A Projection Approach},author={Lesniewski, Andrew},year={2025},eprint={2503.21984},archivePrefix={arXiv},url={https://arxiv.org/abs/2503.21984}}

@article{stokes2020,title={Quantum Natural Gradient},author={Stokes, James and Izaac, Josh and Killoran, Nathan and Carleo, Giuseppe},journal={Quantum},volume={4},pages={269},year={2020},doi={10.22331/q-2020-05-25-269}}

@misc{farhi2014,title={A Quantum Approximate Optimization Algorithm},author={Farhi, Edward and Goldstone, Jeffrey and Gutmann, Sam},year={2014},eprint={1411.4028},archivePrefix={arXiv}}

@misc{lloyd2020,title={Quantum embeddings for machine learning},author={Lloyd, Seth and Schuld, Maria and Ijaz, Aroosa and Izaac, Josh and Killoran, Nathan},year={2020},eprint={2001.03622},archivePrefix={arXiv}}

@article{failde2023,title={Using Differential Evolution to avoid local minima in Variational Quantum Algorithms},author={Fa{\'i}lde, Daniel and Viqueira, Jos{\'e} Daniel and Mussa Juane, Mariamo and G{\'o}mez, Andr{\'e}s},journal={Scientific Reports},volume={13},pages={16230},year={2023},doi={10.1038/s41598-023-43404-3}}

@misc{manifoldsjl,title={Manifolds.jl: Projective space, Sphere, and Symmetric positive definite documentation},author={{JuliaManifolds contributors}},year={2026},note={Accessed 25 October 2026},url={https://juliamanifolds.github.io/Manifolds.jl/stable/manifolds/projectivespace/}}

@article{shan2026,title={A swarm intelligence optimization algorithm on Riemannian manifolds},author={Shan, Min and Halimu, Yeerjiang and Sun, Jun and Fang, Wei and Palade, Vasile},journal={Swarm and Evolutionary Computation},volume={101},pages={102319},year={2026},doi={10.1016/j.swevo.2026.102319}}

@article{piotrowski2026,title={Experimental survey of {L-SHADE} and {SHADE}-based adaptive differential evolution algorithms},author={Piotrowski, Adam P. and Piotrowska, Agnieszka E. and Napiorkowski, Jaroslaw J.},journal={Swarm and Evolutionary Computation},volume={101},pages={102286},year={2026},doi={10.1016/j.swevo.2026.102286}}

@article{pennec2006,title={Intrinsic statistics on {R}iemannian manifolds: Basic tools for geometric measurements},author={Pennec, Xavier},journal={Journal of Mathematical Imaging and Vision},volume={25},number={1},pages={127--154},year={2006},doi={10.1007/s10851-006-6228-4}}

@article{das2011,title={Differential evolution: A survey of the state-of-the-art},author={Das, Swagatam and Suganthan, Ponnuthurai Nagaratnam},journal={IEEE Transactions on Evolutionary Computation},volume={15},number={1},pages={4--31},year={2011},doi={10.1109/TEVC.2010.2059031}}

@article{das2016,title={Recent advances in differential evolution -- An updated survey},author={Das, Swagatam and Mullick, Sankha Subhra and Suganthan, Ponnuthurai N.},journal={Swarm and Evolutionary Computation},volume={27},pages={1--30},year={2016},doi={10.1016/j.swevo.2016.01.004}}

@incollection{zelinka2004,title={{SOMA} -- Self-Organizing Migrating Algorithm},author={Zelinka, Ivan},booktitle={New Optimization Techniques in Engineering},editor={Onwubolu, Godfrey C. and Babu, B. V.},series={Studies in Fuzziness and Soft Computing},volume={141},pages={167--217},publisher={Springer},year={2004}}

@article{skanderova2023,title={Self-organizing migrating algorithm: review, improvements and comparison},author={Skanderova, Lenka},journal={Artificial Intelligence Review},volume={56},number={1},pages={101--172},year={2023},doi={10.1007/s10462-022-10167-8}}

@inproceedings{pluhacek2021,title={Explaining {SOMA}: The relation of stochastic perturbation to population diversity and parameter space coverage},author={Pluhacek, Michal and Kazikova, Anezka and Kadavy, Tomas and Viktorin, Adam and Senkerik, Roman},booktitle={Proceedings of the Genetic and Evolutionary Computation Conference Companion (GECCO '21)},publisher={ACM},year={2021}}

@inproceedings{kadavy2021,title={Self-organizing migrating algorithm with clustering-aided migration and adaptive perturbation vector control},author={Kadavy, Tomas and Pluhacek, Michal and Viktorin, Adam and Senkerik, Roman},booktitle={Proceedings of the Genetic and Evolutionary Computation Conference Companion (GECCO '21)},publisher={ACM},year={2021},doi={10.1145/3449726.3463212}}

@inproceedings{borckmans2010,title={A modified particle swarm optimization algorithm for the best low multilinear rank approximation of higher-order tensors},author={Borckmans, Pierre B. and Ishteva, Mariya and Absil, Pierre-Antoine},booktitle={Swarm Intelligence (ANTS 2010)},series={Lecture Notes in Computer Science},volume={6234},pages={13--23},publisher={Springer},year={2010},doi={10.1007/978-3-642-15461-4_2}}

@article{bonnabel2013,title={Stochastic gradient descent on {R}iemannian manifolds},author={Bonnabel, Silv{\`e}re},journal={IEEE Transactions on Automatic Control},volume={58},number={9},pages={2217--2229},year={2013},doi={10.1109/TAC.2013.2254619}}

@inproceedings{sutton2007,title={Differential evolution and non-separability: Using selective pressure to focus search},author={Sutton, Andrew M. and Lunacek, Monte and Whitley, L. Darrell},booktitle={Proceedings of the 9th Annual Conference on Genetic and Evolutionary Computation (GECCO '07)},pages={1428--1435},publisher={ACM},year={2007},doi={10.1145/1276958.1277221}}

@article{hansen2011,title={Impacts of invariance in search: When {CMA-ES} and {PSO} face ill-conditioned and non-separable problems},author={Hansen, Nikolaus and Ros, Raymond and Mauny, Nikolas and Schoenauer, Marc and Auger, Anne},journal={Applied Soft Computing},volume={11},number={8},pages={5755--5769},year={2011},doi={10.1016/j.asoc.2011.03.001}}

@article{hansen2001,title={Completely derandomized self-adaptation in evolution strategies},author={Hansen, Nikolaus and Ostermeier, Andreas},journal={Evolutionary Computation},volume={9},number={2},pages={159--195},year={2001},doi={10.1162/106365601750190398}}

@inproceedings{tanabe2013,title={Success-history based parameter adaptation for differential evolution},author={Tanabe, Ryoji and Fukunaga, Alex},booktitle={2013 IEEE Congress on Evolutionary Computation},pages={71--78},year={2013},doi={10.1109/CEC.2013.6557555}}

@inproceedings{tanabe2014,title={Improving the search performance of {SHADE} using linear population size reduction},author={Tanabe, Ryoji and Fukunaga, Alex S.},booktitle={2014 IEEE Congress on Evolutionary Computation},pages={1658--1665},year={2014},doi={10.1109/CEC.2014.6900380}}

@article{derrac2011,title={A practical tutorial on the use of nonparametric statistical tests as a methodology for comparing evolutionary and swarm intelligence algorithms},author={Derrac, Joaqu{\'i}n and Garc{\'i}a, Salvador and Molina, Daniel and Herrera, Francisco},journal={Swarm and Evolutionary Computation},volume={1},number={1},pages={3--18},year={2011},doi={10.1016/j.swevo.2011.02.002}}

@article{holm1979,title={A simple sequentially rejective multiple test procedure},author={Holm, Sture},journal={Scandinavian Journal of Statistics},volume={6},number={2},pages={65--70},year={1979}}

@inproceedings{auger2005,title={Performance evaluation of an advanced local search evolutionary algorithm},author={Auger, Anne and Hansen, Nikolaus},booktitle={2005 IEEE Congress on Evolutionary Computation},volume={2},pages={1777--1784},year={2005},doi={10.1109/CEC.2005.1554903}}

@article{hansen2021coco,title={{COCO}: A platform for comparing continuous optimizers in a black-box setting},author={Hansen, Nikolaus and Auger, Anne and Ros, Raymond and Mersmann, Olaf and Tu{\v s}ar, Tea and Brockhoff, Dimo},journal={Optimization Methods and Software},volume={36},number={1},pages={114--144},year={2021},doi={10.1080/10556788.2020.1808977}}

@article{peruzzo2014,title={A variational eigenvalue solver on a photonic quantum processor},author={Peruzzo, Alberto and McClean, Jarrod and Shadbolt, Peter and Yung, Man-Hong and Zhou, Xiao-Qi and Love, Peter J. and Aspuru-Guzik, Al{\'a}n and O'Brien, Jeremy L.},journal={Nature Communications},volume={5},pages={4213},year={2014},doi={10.1038/ncomms5213}}

@article{cerezo2021,title={Variational quantum algorithms},author={Cerezo, M. and Arrasmith, Andrew and Babbush, Ryan and Benjamin, Simon C. and Endo, Suguru and Fujii, Keisuke and McClean, Jarrod R. and Mitarai, Kosuke and Yuan, Xiao and Cincio, Lukasz and Coles, Patrick J.},journal={Nature Reviews Physics},volume={3},pages={625--644},year={2021},doi={10.1038/s42254-021-00348-9}}

@article{kandala2017,title={Hardware-efficient variational quantum eigensolver for small molecules and quantum magnets},author={Kandala, Abhinav and Mezzacapo, Antonio and Temme, Kristan and Takita, Maika and Brink, Markus and Chow, Jerry M. and Gambetta, Jay M.},journal={Nature},volume={549},pages={242--246},year={2017},doi={10.1038/nature23879}}

@article{mcclean2018,title={Barren plateaus in quantum neural network training landscapes},author={McClean, Jarrod R. and Boixo, Sergio and Smelyanskiy, Vadim N. and Babbush, Ryan and Neven, Hartmut},journal={Nature Communications},volume={9},pages={4812},year={2018},doi={10.1038/s41467-018-07090-4}}

@inproceedings{lavrijsen2020,title={Classical optimizers for noisy intermediate-scale quantum devices},author={Lavrijsen, Wim and Tudor, Ana and M{\"u}ller, Juliane and Iancu, Costin and de Jong, Wibe},booktitle={2020 IEEE International Conference on Quantum Computing and Engineering (QCE)},pages={267--277},year={2020},doi={10.1109/QCE49297.2020.00041}}

@article{bonetmonroig2023,title={Performance comparison of optimization methods on variational quantum algorithms},author={Bonet-Monroig, Xavier and Wang, Hao and Vermetten, Diederick and Senjean, Bruno and Moussa, Charles and B{\"a}ck, Thomas and Dunjko, Vedran and O'Brien, Thomas E.},journal={Physical Review A},volume={107},pages={032407},year={2023},doi={10.1103/PhysRevA.107.032407}}

@article{nakanishi2020,title={Sequential minimal optimization for quantum-classical hybrid algorithms},author={Nakanishi, Ken M. and Fujii, Keisuke and Todo, Synge},journal={Physical Review Research},volume={2},pages={043158},year={2020},doi={10.1103/PhysRevResearch.2.043158}}

@article{ostaszewski2021,title={Structure optimization for parameterized quantum circuits},author={Ostaszewski, Mateusz and Grant, Edward and Benedetti, Marcello},journal={Quantum},volume={5},pages={391},year={2021},doi={10.22331/q-2021-01-28-391}}

@article{haug2021,title={Capacity and quantum geometry of parametrized quantum circuits},author={Haug, Tobias and Bharti, Kishor and Kim, M. S.},journal={PRX Quantum},volume={2},pages={040309},year={2021},doi={10.1103/PRXQuantum.2.040309}}

@article{larocca2023,title={Theory of overparametrization in quantum neural networks},author={Larocca, Mart{\'i}n and Ju, Nathan and Garc{\'i}a-Mart{\'i}n, Diego and Coles, Patrick J. and Cerezo, M.},journal={Nature Computational Science},volume={3},pages={542--551},year={2023},doi={10.1038/s43588-023-00467-6}}

@misc{novak2026qaoa,title={Adaptive Differential Evolution and multistart search for noisy {QAOA} optimization},author={Nov{\'a}k, Vojt{\v e}ch and Zelinka, Ivan and Das, Swagatam and Beseda, Martin},year={2026},note={arXiv:2609.23180}}

@misc{novak2026isoma,title={Adaptive rotation for {iSOMA}: Geometry, benchmarking, and noise robustness in variational quantum objectives},author={Nov{\'a}k, Vojt{\v e}ch and Zelinka, Ivan},year={2026},note={arXiv:2609.37193}}

@misc{novak2026geom,title={Optimization geometry of {QAOA} and variational quantum algorithms},author={Nov{\'a}k, Vojt{\v e}ch and Zelinka, Ivan and Ill{\'e}sov{\'a}, Silvie and Das, Swagatam and Beseda, Martin},year={2026},note={arXiv:2610.05524}}

@misc{zenodo2026,title={Reproducibility package: Geometry, masking and selection in {D}ifferential {E}volution and {SOMA} on quantum state manifolds},author={Nov{\'a}k, Vojt{\v e}ch and Zelinka, Ivan},year={2026},publisher={Zenodo},note={\url{https://doi.org/10.5281/zenodo.24516254}}}

@article{novak2026noisy,title={Optimization strategies for variational quantum algorithms in noisy landscapes},author={Nov{\'a}k, Vojt{\v e}ch and Zelinka, Ivan and Sn{\'a}{\v s}el, V{\'a}clav},journal={Evolutionary Intelligence},volume={19},number={5},pages={142},year={2026}}

@misc{illesova2025vha,title={Numerical optimization strategies for the variational {H}amiltonian ansatz in noisy quantum environments},author={Ill{\'e}sov{\'a}, Silvie and Nov{\'a}k, Vojt{\v e}ch and Bezd{\v e}k, Tom{\'a}{\v s} and Possel, Christian and Beseda, Martin},year={2025},note={arXiv:2505.22398}}

@article{illesova2025bench,
  title = {Statistical benchmarking of optimization methods for a variational quantum eigensolver under quantum noise},
  author = {Illésová, Silvie and Bezdĕk, Tomáš and Novák, Vojtĕch and Senjean, Bruno and Beseda, Martin},
  journal = {Phys. Rev. A},
  pages = {},
  year = {2026},
  month = {Nov},
  publisher = {American Physical Society},
  doi = {10.1103/5clr-mhwn},
  url = {https://link.aps.org/doi/10.1103/5clr-mhwn}
}

@article{novak2026cec,title={A longitudinal analysis of the {CEC} single-objective competitions (2010--2024) and implications for variational quantum optimization},author={Nov{\'a}k, Vojt{\v e}ch and Bezd{\v e}k, Tom{\'a}{\v s} and Zelinka, Ivan and Das, Swagatam and Beseda, Martin},journal={Swarm and Evolutionary Computation},volume={107},pages={102469},year={2026}}

@misc{bezdek2025,title={Classical optimization strategies for variational quantum algorithms: A systematic study of noise effects and parameter efficiency},author={Bezd{\v e}k, Tom{\'a}{\v s} and Yuan, H. and Nov{\'a}k, Vojt{\v e}ch and Ill{\'e}sov{\'a}, Silvie and Beseda, Martin},year={2025},note={arXiv:2511.09314}}

@misc{novak2025reliable,title={Reliable optimization under noise in quantum variational algorithms},author={Nov{\'a}k, Vojt{\v e}ch and Ill{\'e}sov{\'a}, Silvie and Bezd{\v e}k, Tom{\'a}{\v s} and Zelinka, Ivan and Beseda, Martin},year={2025},note={arXiv:2511.08289}}

@article{liu2016mde,title={Manifold Differential Evolution ({MDE}): A Global Optimization Method for Geodesic Centroidal {V}oronoi Tessellations on Meshes},author={Liu, Yong-Jin and Xu, Chun-Xu and Yi, Ran and Fan, Dian and He, Ying},journal={ACM Transactions on Graphics},volume={35},number={6},pages={243},year={2016},doi={10.1145/2980179.2982424}}
\end{document}